\documentstyle[12pt,aasms4]{article}

\received{Dec 3, 1997}
\accepted{Feb 27, 1998}
\slugcomment{To appear in {\it The Astrophysical Journal}, July 20, 1998}

\lefthead{Allen \& Kronberg}
\righthead{Radio Spectra}

\newcommand{\etal}{{\it{et al. }}}
\newcommand{\nd}{\nodata}
\newcommand{\HII}{\ion{H}{2} }
\newcommand{\NeII}{[\ion{Ne}{2}] }

\begin{document}

\title{Detailed Radio Spectra of Selected Compact\\ 
Sources in the Nucleus of M82}
\author{Michael L. Allen \& Philipp P. Kronberg}
\affil{Dept. of Astronomy\\University of Toronto\\
60 St. George Street\\Toronto, Ontario\\M5S 3H8 CANADA}

\begin{abstract}
We have determined detailed radio spectra for 26 compact sources in the 
starburst nucleus of M82, between
$\lambda\lambda$ 74 and 1.3 cm.  Seventeen show low-frequency turnovers.  One
other has a thermal emission spectrum, and we identify it as an \HII 
region.  The low frequency turnovers are due to absorption by
interstellar thermal gas in M82.

New information on the AGN candidate, 44.01+595, shows it to have a 
non-thermal falling powerlaw spectrum at the highest frequencies, and that 
it is strongly absorbed below 2 GHz.

We derive large magnetic fields in the supernova remnants, of
order 1-2 $(1+k)^{2/7} \phi^{-2/7}$ milliGauss, hence large pressures in the 
sources suggest that the brightest ones are either expanding or are
strongly confined by a dense interstellar medium.
From the largest source in our sample, we derive a supernova rate of 
0.016 yr$^{-1}$.
\end{abstract}

\keywords{galaxies: individual (M82) --- galaxies: starburst --- 
radio continuum: galaxies --- supernova remnants}

\section{Introduction}

Supernovae (SNe) and supernova remnants (SNR's) are thought to be the main
``drivers'' of the starburst phenomenon (\cite{kw75}).
Detailed radio observations for supernovae and supernova remnants 
are few, due to the fact that only a small subset of the 
extragalactic SNe discovered each year produce detectable radio emission. 
Those events 
for which monitoring data is available (e.g. \cite{weiler88}) have revealed
considerable information regarding the evolution of the properties of the 
expanding shock wave and its emission processes.  In intense starbursts like
M82, where a significant population of massive stars has had sufficient time
to evolve to the SN stage, 
we have the unique opportunity to study a collection of SNR's of similar
age and origin, and their interaction with the surrounding environment.

The observations are briefly outlined in Section 2.  We discuss the 
technique used to fit spectral models in Section 3.  Then in Section 4
we discuss the physical implications of our modelling. 
Throughout the paper, we assume that the distance to M82 is 3.63 Mpc
(\cite{freedman}, \cite{tammann}), so that 1$\farcs$0 corresponds to 17.6 pc.

\section{Observations}

We have used observations of M82 at $\lambda\lambda$ 20, 6, 3.6, 2, and 1.3 cm.
To supplement these observations, we have added published flux densities 
at $\lambda\lambda$ 74 and 18 cm (408 and 1666 MHz) for individual radio point 
sources as measured with the MERLIN interferometer by \cite{unger84} and
\cite{wills97}.

The $\lambda\lambda$ 2 and 1.3 cm data were first presented in 
\cite{golla96} and are composed of 
VLA\footnote{The NRAO is operated by Associated Universities, Inc., under 
cooperative agreement with the National Science Foundation.}
data collected at all four configurations, including data from the flux 
monitoring program 
of Kronberg \& Sramek (see e.g. Kronberg \& Sramek 1992) between 1983 and 1992.
The $\lambda$ 20 cm data is from two A-array observations (\cite{golla96}).
The $\lambda$ 6 cm data is a combination of VLA data from its A-, C- and 
D-configurations.
The $\lambda$ 3.6 cm data were collected 
in 1994-5 in the VLA's A-, B- and C-configurations, and were combined with a 
previous observation in the D-configuration, first presented in 
\cite{reuter93}.

\section{Data Reduction}

\subsection{Source Fluxes}

The combined datasets at $\lambda\lambda$20, 6, 3.6, 2 and 1.3 cm were mapped 
and CLEAN-deconvolved using the NRAO's AIPS imaging software.  To eliminate 
the bright background 
emission and concentrate on the point sources, we restricted the lower 
$uv$-baseline range so that no structures spatially larger than 4$\arcsec$ 
were present.  This range proved most effective - 
restricting the data to smaller structures led to point
spread functions that had large negative sidelobes, that in turn led to 
deconvolution 
problems; allowing larger structures led to a non-zero background that 
tended to cause problems with the determination of the integrated fluxes of
the point sources.
The complete list of images used is given in Table 1.

A full description of the data reduction of the combined VLA datasets will 
be given in a forthcoming paper, where we will also present a study of 
the diffuse radio emission defining the nuclear region of M82.

\placetable{tbl-1}

The datasets were   mapped at 0$\farcs$30  and 0$\farcs$20 resolution.
These beamsizes  were chosen for proper flux  density scaling with the
lower frequency  data of Unger \etal,  which has a FWHM of 0$\farcs$25
at 18cm.  At  the same time  we  wanted to guard against  missing flux
from possible partial resolution of the radio sources.  To each source
we fitted  2-D  gaussian profiles  using the AIPS  task IMFIT.   For a
given wavelength, we then compared the results of  the fitting at each
resolution against each other, and against the result of adding up the
flux in each pixel of the radio source (AIPS task TVSTAT).

For this latter pixel-adding method, we only included those pixels of
the radio source that contained emission above the background local
to the source.  This background was determined by eye, by examining 
profiles of slices through the sources.  Fluxes determined in this way 
served as a quick check on the gaussian fits, and were not used as
data points in any of our spectral fits.

In general, the values for the integrated fluxes determined using each method
agreed well with each other, although the pixel-added TVSTAT fluxes were 
generally smaller.  The flux values quoted in Table 2 are the weighted 
average of the gaussian-fitted fluxes obtained at both resolutions, except
for sources that were partially resolved at 0$\farcs$20, for which we used
only the 0$\farcs$30 results.  These latter were unresolved in all cases.
The errors given for each fit are the larger of the errors calculated by 
IMFIT, and the 3-$\sigma$ level on each map.
For the sources in M82, these resolutions are ideal:  they are sufficiently
high to isolate the sources from the background emission, and yet not 
so high as to resolve the sources out.

The best resolution obtainable for the $\lambda$ 20 cm data was a 
1$\farcs$15$\times$0$\farcs$92 beam in position angle 38$\fdg$6 (ccw from 
north), which matches the resolution of the $\lambda$ 74 cm data.  We chose 
only those point sources that were clearly separated from any neighbours.

We compared our $\lambda$ 3.6 cm results against those of \cite{huang93}, who 
had a resolution of 0$\farcs$182.  Our fluxes are systematically greater than 
theirs.  On average, the values of \cite{huang93} are only 66\% of our values.
We speculate that the difference is due to a combination of 
(1) missing flux on their part due to partial resolution of the sources, and
(2) possible inclusion on our part of contaminating larger-scale flux 
along the same line of sight to the source.  Since we intentionally
excluded structures larger than 4$\arcsec$, this presumed 
large-scale flux would be of very low surface-brightness and would be 
physically unrelated to the compact radio sources.  The values listed in this
publication are consistent with a power-law interpretation of the 
high-frequency shape of the spectra of the point sources.

\placetable{tbl-2}

The complete results for the fluxes of 25 radio point sources are given in 
Table 2 (we treat 41.9+58 separately, in Table 3).  
Of the 19 sources listed by \cite{unger84},
we found that one of their sources (47.7+62) had no corresponding
feature in our maps, and two of their sources could not be identified 
uniquely with a particular source at our wavebands.  On the other hand, 
we were able to identify every source listed by \cite{wills97}.

\paragraph{The source 41.9+58}
To correct for the decaying flux of 41.9+58, we used the variability results 
from 
\cite{ks92} to determine that the approximate decay rate is 8.8\%/year. We
applied this rate to the fluxes measured by Wills \etal.  Our chosen epoch for
the maps is May 1993 (``1993.4'').  
The flux values used for the spectral fit are listed in Table 3.

We note that the corrected epoch-1984 408 MHz flux is 72$\pm$8 mJy,
whereas the corrected epoch-1997 flux is 117$\pm$8 mJy.  
This discrepancy
can be understood by realizing that for this source, \cite{kbs85}
discovered that the turnover frequency is evolving over time, as is the
spectral index at all frequencies.  In particular, the source is becoming 
optically thin at progressively lower frequencies.
This effect has also been seen in other young radio supernovae
(\cite{weiler86}, \cite{weiler88}).
The measured decay rate of 8.8\%/year is derived from $\lambda$ 6 cm (4.8 GHz)
data. Therefore our flux corrections are most accurate within the
powerlaw regime of the spectrum, i.e. the higher frequencies, where the
spectral index has been evolving slowly.  Our decay estimate is least
certain at the lower frequencies where the emission is  
optically thick.  Overall we are probably underestimating the 
errors associated with the flux of 41.9+58.

\placetable{tbl-3}

\subsection{Spectral Fits}

To each point source, we fitted two composite models to the 
emission spectrum: (1) a model of
synchrotron and thermal (\HII) emission, with a thermally absorbing screen
located between the source and the observer, and (2) a co-spatial mix of 
synchrotron and thermally-absorbing gas.
These two models are represented by the equations (cf. \cite{pach77}):
\begin{eqnarray}
I(\nu) & = & I_{sy} \nu^{\alpha} e^{-(\nu/\nu_{\tau=1})^{-2.1}}
             + I_{th} \left( \frac{\nu}{\nu_{\tau=1}} \right)^2
             \left[ 1 - e^{-(\nu/\nu_{\tau=1})^{-2.1}} \right] 
             \;\;\;\;(screen)\\
I(\nu) & = & \left[ I_{sy} \nu^{2.1+\alpha} + I_{th} 
             \left( \frac{\nu}{\nu_{\tau=1}} \right)^2 \right]
             \left[ 1 - e^{-(\nu/\nu_{\tau=1})^{-2.1}} \right]
             \;\;\;\;(mixed)
\end{eqnarray}
In the above expressions, $I_{sy}$ and $I_{th}$ are related in a simple way
to the optically thin synchrotron and thermal fluxes respectively, $\alpha$ 
is the (optically thin) non-thermal spectral index (defined so that 
$S \propto \nu^{+\alpha}$), and $\nu_{\tau=1}$ is the frequency at 
which the optical depth to thermal absorption becomes unity.

Since the number of data points and free parameters are comparable, standard
least-squares fitting procedures were inadequate for finding the absolute 
minimum of the chi squared hypersurface.  Therefore, we wrote out the $\chi^2$
equations and their derivatives in full.  
It was discovered that two of the 
parameters, ($I_{sy}$ and $I_{th}$) could be expressed completely in terms
of the other two ($\alpha$ and $\nu_{\tau=1}$).  A grid over reasonable
values of these two latter parameters was constructed, 
and a value for $\chi^2$
at each point was generated.  We chose the set of ($\alpha$, $\nu_{\tau=1}$) 
that yielded the lowest value for $\chi^2$. The spectral ``best'' fits 
are included in Tables 4, 5 and 6, and are plotted in Figure 1.

In Table 4, we list the best fit parameters for those sources
that have no firm evidence for absorption.  Tables 5 and 6 list the
best fit parameters for absorbed sources - for Table 5 we have used a
screen model fit (equation 1), and for Table 6 we have used a 
mixed model fit (equation 2).  Only one source had a very uncertain
spectral shape, 47.37+680.  
For this source we plotted both the power-law fit and a fit for 
the screen model; for the mixed model fit no sensible solution was reached.

To evaluate the reliability of our fits, we generated test data, 
and examined slices through the $\chi^2$ surface.  The surfaces 
were irregularly shaped, with a pronounced absolute minimum at the location
expected for the test data.  By varying the spectral
index $\alpha$ between -0.11 and -2.20 in increments of 0.01, and the optical
thickness parameter $\nu_{\tau=1}$ from 0.10 to 4.00 GHz in increments of
0.01 GHz, we were able to unambiguously identify the best fit solution
within an error equal to the size of the grid increments. 

A separate test on each source was applied to see if the spectrum could
be purely thermal ($S \propto \nu^{-0.1}$).

\placetable{tbl-4}

\placetable{tbl-5}

\placetable{tbl-6}

\section{Results}

\subsection{Steep-spectrum Sources}

\paragraph{Magnetic field, associated pressures}
Fully 19 of the 26 sources show optically thin, non-thermal spectral indices 
of $\alpha$ $<$ -0.49.

For those sources we calculated the 
equipartition magnetic field strength (cf. \cite{pach2}). This value is derived
from the calculated minimum energy in particles and fields required to 
produce the observed radio signature, within an estimated radiating volume.  
The values scale with the
undetermined parameters $k$, which is the ratio of the relativistic proton to 
electron energy densities, and $\phi$, which 
is the fraction of the source's volume occupied by relativistic particles and 
fields (refer to the notes in Tables 4, 5 \& 6).  
From direct cosmic ray observations near the solar system, it is 
thought that $k$ is between 
40-100 in the Milky Way (see e.g. \cite{morfill76}).

We assumed that the intrinsic spectral index for each source was constant 
from 0.01 to 100 GHz. In most of the sources, the equipartition magnetic 
field is about 1-2 mGauss.  Bright, compact radio sources in other galaxies 
have similar inferred field strengths (\cite{condon91}).  These values are 
higher than but comparable to the inferred ISM magnetic field strength for M82
of $\sim$ 50-100 $\mu$G (\cite{rieke80}, \cite{klein88}, \cite{condon92}), 
and are much higher than the measured value for the magnetic field in
our Galaxy, $\sim$ 3 $\mu$Gauss (\cite{spitzer}). 

The corresponding minimum pressures of the synchrotron plasma
within the point sources have been calculated using equation (7.15) of
\cite{pach2}. They are of order $\sim$ 10$^{-8}$ dynes cm$^{-2}$.
These values are derived assuming that the sources have particles and fields
distributed homogeneously throughout a spherical volume, and hence represent
lower limits.
  
The total pressure of the ISM in the nucleus of M82 has been estimated by 
\cite{mccarthy87}.  They derive values $\sim$ 10$^{-8}$ $\rightarrow$
 10$^{-8.5}$ dynes cm$^2$.
These values would suggest that the brighter point sources (at least)
in this survey are expanding into a lower-pressure medium.
All of these values are far higher than the pressure of the ISM of the Galaxy,
which is $\sim$ 10$^{-13}$ dynes cm$^{-2}$ (\cite{spitzer}).
This evidence supports the theory that supernovae add mechanical and
magnetic energy into the ISM of starburst galaxies.

The spectral fits for the steep-spectrum sources ($\alpha$ $<$ -0.49)
show very low levels of free-free emission.  These values
are generally so low as to have a negligible effect on the actual shape of 
the spectra.  The exceptions are sources 43.31+591, 46.52+638, 46.56+738, 
and 46.70+670.

\paragraph{Low frequency turnovers}
The most striking feature of the spectra is that 17 of the 20 steep-spectrum 
sources (and one flat-spectrum source, see below) show a low-frequency
turnover. We will consider the nature of this phenomenon in this section.
The discussion applies to homogeneous sources.

A turnover at the lower end of a synchrotron spectrum can result from many
situations:  the synchrotron source can become optically thick due to 
self-absorption, the injected relativistic electrons can have a low-energy
cutoff, plasma effects, or absorption by intervening \HII.
We will show that intervening \HII is the most likely cause.

Optically thick synchrotron emission has a spectral index 
$\alpha = + \slantfrac{5}{2}$, where $I(\nu) \propto \nu^{\alpha}$.
Synchrotron self-absorption becomes important for sources with a brightness
temperature $T_b > m_e c^2/k_B \sim 10^{10}$ K.  The brightest 
source in our sample (41.9+58) has $T_b \approx 10^7$ K at 1.4 GHz, well 
below the synchrotron self-absorption threshold, so we can
discount self-absorption because the surface brightnesses are too low.

If the synchrotron radiation is emitted in a plasma, then the presence of
thermal electrons will decrease the emission efficiency of the 
relativistic electrons, called the Tsytovich-Razin effect (\cite{pach2},
\cite{pach77}).  This effect becomes important at frequencies 
$\nu \lesssim 20 N_e/B$ (cgs).  For our derived $B \sim 1$ mG, this effect
becomes noticeable at 1 GHz for an electron number density of 
$N_e \sim 5.0 \times 10^4$ cm$^{-3}$.  At such high densities, however,
free-free absorption is a more important effect.  The free-free optical depth 
is
\begin{equation}
        \tau_{ff} \sim 3.3 \times 10^{-7} 
        \left( \frac{N_e}{cm^{-3}} \right)^2
        \left( \frac{T_e}{10^4 K} \right)^{-1.35} 
        \left( \frac{\nu}{GHz} \right)^{-2.1}
        \left( \frac{s}{pc} \right).
\end{equation}
At N$_e$ = 5 $\times$ 10$^4$ cm$^{-3}$, T$_e$ = 10$^4$ K, $\nu$ = 1 GHz, 
and for a typical M82 SNR of radius s = 0.8 pc (0.05\arcsec), the free-free
optical depth is 660!  We can thus discount a Tsytovich-Razin-like cutoff
to explain the low-frequency shape of the spectra.

If the source is optically thin at all frequencies, then the spectrum will 
have a spectral index $\alpha = + \slantfrac{1}{3}$ at frequencies below the 
lower cutoff frequency. This slope is characteristic of the emission
from a single synchrotron electron at frequencies below its critical frequency.
Two sources (40.68+550 and 44.01+595) have spectral indices steeper than
$ + \slantfrac{1}{3}$ at the lowest measured frequencies, which requires
some absorption mechanism.  Since the spectral properties of the compact
sources are similar, it is most probable that the mechanism responsible 
for the low frequency turnover is interstellar, and not a property 
intrinsic to each individual source.  Therefore we did not test for 
an intrinsic lower cutoff frequency.
It would be clearly desirable to obtain even lower frequency
measurements, down to 150 MHz, in order to probe the spectra most fully.

Rather, we conclude that free-free absorption is the most important effect.
Our spectra were fit by models involving absorbing \HII gas, located either
(1) in an intervening screen located between the source and the telescope, or 
(2) mixed with the synchrotron-emitting particles.  From the data, it is not 
possible to dismiss either situation on statistical grounds alone. 
To better understand the nature of this absorbing gas, we need to examine
other tracers of \HII in the nucleus of M82.

As remarked by \cite{aandl95} in an analysis of their \NeII data, 
the SNR's are found outside of regions of highest \NeII brightness, i.e. 
ionized gas.  Consistent with this observation, no correlation is found 
between the positions of the SNR and the radio recombination line (RRL)
images of \cite{seaquist96}.  These data trace \HII gas that
is located throughout the nuclear region of the galaxy, and perhaps 
concentrated in a ring, as proposed by \cite{aandl95}.

We will try to estimate whether there is enough of this ambient gas to cause
the absorption effects we see here.  \cite{seaquist96} state that the density
of thermal electrons is between 10-100 cm$^{-3}$.  Since the SNR's avoid the 
brightest regions of RRL emission, we assume an average value of 50 cm$^{-3}$
for interstellar thermal electrons.
At a temperature $T_e = 10^4$ K, the SNR emission at 1 GHz would have to
travel a distance of over 1 kpc to be extinguished to an optical depth of 1.
This size represents the entire extent of the radio nucleus of M82, and would
indicate that all the ``extinguished'' SNR's (i.e. the majority of them) were 
located at the far side of the nuclear region.  

It is more likely that the SNR's are distributed at a variety of depths into
the nuclear region. We therefore need to test the optical depth of more dense
clouds.  For an \HII cloud with $N_e = 10^4$ cm$^{-3}$ and $T_e = 10^4$ K, 
at 1 GHz radio photons need travel less than 0.03 pc to be extinguished to 
an optical depth of 1.  The RRL data of \cite{seaquist96}
suggest that regions with high electron densities do exist, and that they
have a very low filling factor.

We support the scenario favoured by many authors, that the ISM in M82 is 
extremely non-uniform, in which the energy from stellar winds and SNe 
completely
dominates the state of the gas. Only extremely dense clouds have survived
the disruptive effects of these processes.  These clouds cause the absorption
seen at the longer wavelengths in the compact sources, either because they
lie along the same line of sight to the source, or the absorbing gas is
associated with the source itself.

\subsubsection{Notes on Individual Sources}

\paragraph{44.01+595}
Spectrum fitting of this source by \cite{wills97} suggested that it is
unusual in that it has a rising spectrum towards higher frequencies.
They, and other authors (e.g. \cite{seaquist97}) consider it to be an
AGN candidate.
Our data shows the opposite - the source has a falling, power-law spectrum
between $\lambda\lambda$ 3.6 and 1.3 cm.  The source shows the strongest
absorption in this survey, being optically thick below 2.01 GHz.
The spectrum is therefore rising below 2 GHz, and this fact is probably the 
cause of the previous confusion concerning the shape of the spectrum.
The radio continuum spectrum alone is not a good discriminator for AGN
candidates.  We feel that the evidence to date concerning this source
raises intriguing questions as to its identity.  
A detailed VLBI study will likely be necessary to determine the exact
properties of this source.

%%[suggest some kind of discriminating AGN observation
%%mention that the source is compact, no unusual variability,
%%slightly resolved in VLBI so very young SNR if so, peak of molecular gas]

\paragraph{47.37+680}  This source is not included in the histogram of spectral
indices (Figure 2).  Both
powerlaw and curve fits seem to work, but the curve fit predicts fainter
high-frequency emission, hence fits the sense of the data better.

This source has the shape of a ring or shell in our highest
resolution maps, and therefore its integrated flux was determined by
adding up the flux on a pixel-by-pixel basis using the AIPS task TVSTAT.
We show an X-band map of the source in Figure 4, at a resolution of 
0$\farcs$23.  The source is roughly
circular, with a diameter of about 0$\farcs$91, or about 16 pc. If the remnant
has been expanding undecelerated, at a rate V (in km s$^{-1}$), then its
age would be $\sim$ 3100 $\left( \frac{V}{5000 \;\;km/s} \right)^{-1}$ years.

From its large size, this remnant would be judged the oldest one in our sample.
Assuming that the $\sim$50 other sources are radio SNR, we estimate a
supernova rate of $\sim$0.016 $\left( \frac{V}{5000 \;\;km/s} \right)$
yr$^{-1}$, or, assuming V=5000 km s$^{-1}$, one supernova explosion
every 62 years.  This estimate is lower than those found by other authors,
however, increasingly sensitive surveys have increased the number of known
compact sources, and hence our value is probably a lower limit.

\subsection{Flat-spectrum Sources}

We have designated five sources as flat spectrum 
sources.  We identify source 42.2+590 as an \HII region, in support of earlier
evidence presented by \cite{muxlow94}.
We discuss each source individually, below.

\paragraph{39.11+573}
This source has a best-fit spectral index of $\alpha=-0.38$.
In the 0$\farcs$050 beam of \cite{muxlow94}, it was resolved into a
shell-like structure.  It is almost certainly a SNR.

\paragraph{42.21+590}
Source 42.21+590 has many other properties typical of giant \HII regions.  It 
is the site of maser activity of OH (\cite{wfg84}) and H$_2$O molecules
(\cite{baudry96}).
It is also located near an emission peak at $\lambda$ 3.3 mm (92 GHz), 
at which wavelength the emission is thought to be predominantly free-free
(see \cite{ck91}).

We compared the source position with an image of the 12.8 $\mu$m line of 
\NeII (\cite{aandl95}) given that this line is an ideal tracer of thermal gas.
The source is indeed located within a region of high \NeII brightness.
This same \NeII emission region also encompasses the bright, 
confused region centred near RA 09$^h$ 51$^m$ 42$\fs$5, 
Dec 69$\arcdeg$ 54$\arcmin$ 58$\arcsec$ (for a radio-\NeII overlay, see
\cite{golla96}).  In our radio maps, the two regions appear to be unconnected,
but they may form part of a more massive \HII complex.

The source 42.21+590 is partially resolved in the 0$\farcs$050 beam at 5 GHz 
of \cite{muxlow94}, where they cite a ``largest angular size'' of 
0$\farcs$275 based on the diameter of their 3-$\sigma$ contour.  Our spectral 
fit predicts an optically thick flux at 1 GHz of 0.877 mJy, and assuming that 
the region is spherical, we derive an electron temperature of 2800 K. Using 
equation (2) of \cite{condon92}, the predicted production rate of ionizing 
photons is $N_{uv}$ $\geq$ 2.43$\times$10$^{50}$ $s^{-1}$, which is the 
equivalent of 108 ZAMS O8 stars (\cite{panagia73}). 
We summarize these properties in Table 7.

\placetable{tbl-7}

\paragraph{39.40+561, 45.48+648, 45.74+652}
These three sources cannot be unambiguously identified as either radio
SNR or \HII regions.  They each have a flat spectral index.  Their
brightness temperatures fall in the range $10^4 \rightarrow 10^5$ K. In
our highest resolution maps they are marginally resolved as a brightly
peaked structure with some extended emission.  There is no evidence of
any ring-like morphology. The latter two sources fall in a region of bright, 
clumpy background emission, and there are several nearby point sources 
discernible in our highest-resolution maps.  The morphology of 
45.74+652 is peculiar in that it has a ``tail'' of emission 
extending $\sim$ 1$\farcs$0 to the north-west of the source.

In a separate study, we are undertaking a detailed examination of the 
large-spatial-scale, diffuse emission in M82 in order to better understand 
the relative importance of the thermal from the non-thermal
emission.  The nature of the emission of these larger complexes of emission
must be understood better before we can properly examine the spectra of
those sources imbedded within them.  The results of this study will be 
presented in a later paper.

\section{Conclusions}

From our study of 26 relatively bright compact radio sources in the nucleus 
of M82, we can draw the following conclusions:

1. Of the 26 sources, one (42.21+590) is an \HII region, 22 are most likely
young radio supernovae or supernova remnants, and three further sources remain 
of uncertain nature.

2. High magnetic field strengths and pressures are derived for the
sources we identify as SNR's.  The brighter ones do not appear to be in 
pressure equilibrium with the surrounding medium, and are probably expanding,
or are strongly confined by the high pressure ISM.
The source of the high nuclear pressure and halo winds are most likely
the SNR's.

3. Sources that show a low-frequency turnover, do so as a result of absorption
by ionized gas.  This ionized gas may be intimately associated with the 
source, or may be in clouds located along the line of sight to the sources.  
The gas is most likely in the form of clumpy, high density clouds.

4. The resolved ring-like source in our survey is the oldest.
We derive a supernova rate in M82 of $\sim$ 0.016 
$\left( \frac{V}{5000 \;\;km/s} \right)$yr$^{-1}$.  This value
is a lower limit.

\acknowledgements

MLA wishes to thank Chuck Shepherd for useful discussions involving
statistics and fitting procedures.   
We thank Richard A. Sramek for the use of VLA data.  It is a pleasure to 
thank the Director and staff of the NRAO for their help and generous
allotment of time on the VLA.  The NRAO is operated by Associated 
Universities Inc., under cooperative agreement with the National Science
Foundation.
This research has made use of the NASA/IPAC Extragalactic Database (NED)   
which is operated by the Jet Propulsion Laboratory, California Institute   
of Technology, under contract with the National Aeronautics and Space 
Administration.

\clearpage

%% TABLE 1

\begin{deluxetable}{rrr}
\small
\tablewidth{0pt}
\tablecaption{Maps and Associated noise \label{tbl-1}}
\tablehead{
\colhead{Wavelength} & \colhead{Beam FWHM} & \colhead{RMS noise} \nl
\colhead{(cm)}       & \colhead{(arcsec)}  & \colhead{$\mu$Jy/beam}
}
\startdata
20.0 & 1.147   & 109.8  \nl
6.0  & 0.313   &  45.4  \nl
3.6  & 0.20    &  10.1  \nl
3.6  & 0.30    &  13.3  \nl
2.0  & 0.20    & 117.0  \nl
2.0  & 0.30    &  90.7  \nl
1.3  & 0.20    &  87.8  \nl
1.3  & 0.30    &  77.5  \nl
\enddata
\end{deluxetable}

\clearpage

%% TABLE 2

\begin{deluxetable}{crrrrrrr}
\scriptsize
\tablewidth{0pt}
\tablecaption{Integrated fluxes at the wavelengths measured \label{tbl-2}}
\tablehead{\colhead{Source name\tablenotemark{a}} & 
\colhead{74 cm} & 
\colhead{20 cm\tablenotemark{b}} &
\colhead{18 cm} & 
\colhead{6 cm} & 
\colhead{3.6 cm} &
\colhead{2.0 cm} &
\colhead{1.3 cm}
}
\startdata
39.11+573 &    9.1(2.3)    & 7.92(0.37) & 4.0(1) & 4.59(0.15)& 
        3.77(0.10) & 3.78(0.30) & 1.48(0.27) \nl
39.40+561 &    4.3(1.7)    & \nd & \nd & 2.05(0.15) & 
        2.02(0.10) & 1.10(0.73) & 0.63(0.33) \nl
39.64+534 &    2.6(1.3)    & 2.23(0.78) & \nd & 1.58(0.15) & 
        0.96(0.10) & 0.28(0.30) & 0.41(0.27) \nl
39.77+569 &    2.7(1.3)    & 1.77(0.30) & \nd & 1.06(0.15) & 
        0.75(0.11) & 0.20(0.30) & 0.34(0.27) \nl
40.32+551 &    3.9(2.2)    & \nd & \nd & 0.95(0.15) & 
        0.73(0.10) & 0.37(0.51) & 0.47(0.33) \nl
40.68+550 &   10.9(1.6)    &15.37(1.18) &10.0(1) & 7.31(0.35)& 
        6.10(0.10) & 4.13(0.30) & 2.64(0.27) \nl
41.31+596 & $<$12\phm{(4)} & 4.15(0.33) & 3.0(1) & 3.45(0.15)&
        2.81(0.10) & 1.88(0.34) & 1.29(0.27) \nl
42.21+590 & $<$12\phm{(4)} & \nodata    & 2.5(1) & 4.07(0.50)&
        4.31(0.29) & 4.45(0.30) & 4.12(0.27) \nl
42.53+619 &   11.1(2.4)    &13.14(1.04) & \nd & 1.64(0.15) & 
        1.52(0.50) & 0.25(0.30) & $<$0.27 \nl
42.67+556 &    2.7(1.6)    & \nd & \nd & 1.67(0.15) & 
        1.23(0.10) & 0.81(0.30) & 0.53(0.35) \nl
43.19+583 &    7.9(2.3)    &11.65(0.58) & 8.5(1) & 5.15(0.17)& 
        3.82(0.15) & 2.44(0.30) & 1.77(0.27) \nl
43.31+591 &   37.3(4.5)    &24.69(0.58) &19.5(1) &10.32(0.23)& 
        7.95(0.10) & 5.49(0.30) & 4.09(0.40) \nl
44.01+595 & $<$12\phm{(4)} & 8.88(0.93) & 9.5(1) &21.79(0.89)& 
        20.18(0.10)& 15.79(0.30) & 12.48(0.27) \nl
44.28+592 & $<$12\phm{(4)} & 6.98(0.98) & 3.0(1) & 2.39(0.15)&
        2.19(0.10) &  1.20(0.30) &  0.60(0.27) \nl
44.52+581 &    1.6(1.5)    & 5.15(0.30) & 2.5(1) & 2.63(0.15)& 
        1.99(0.10) & 1.90(0.44) & 0.70(0.27) \nl
45.18+612 &   18.8(2.1)    &20.92(0.32) &12.5(1) & 7.91(0.34)& 
        5.72(0.10) & 3.85(0.30) & 2.53(0.27) \nl
45.25+651 & $<$12\phm{(4)} &  \nodata     & 5.0(1) & 1.79(0.15)& 
        1.58(0.13) & 0.32(0.32) & $<$0.27 \nl
45.43+673 &    7.7(0.8)    &  \nodata   & 2.0(1) & 2.01(0.15)& 
        1.39(0.10) & 0.83(0.30) & 0.52(0.27) \nl
45.48+648 &    2.3(1.3)    & \nd & \nd & 1.32(0.15) & 
        1.16(0.10) & 1.46(0.41) & $<$0.27 \nl
45.74+652 &    8.0(1.9)    &  \nodata   & 2.5(1) & 3.40(0.28)& 
        3.48(0.48) & 3.23(0.50) & 1.33(0.58) \nl
45.86+640 &    3.7(1.9)    & 3.03(0.33) & \nd & 1.90(0.15) & 
        1.31(0.10) & 0.85(0.35) & 0.51(0.38) \nl
46.52+638 &    4.4(1.8)    & 7.18(0.30) & 3.5(1) & 2.74(0.19)& 
        2.08(0.10) & 1.53(0.30) & 1.58(0.27) \nl
46.56+738 &    3.7(0.5)    & 2.65(0.33) & \nd & 1.00(0.15) & 
        0.72(0.10) & 0.59(0.30) & 0.32(0.27) \nl
46.70+670 &    3.9(1.5)    & 4.17(0.45) & 3.0(1) & 1.97(0.15)& 
        1.61(0.10) & 1.00(0.30) & 1.01(0.27) \nl
47.37+680 &    3.8(0.9)    & 3.01(0.33) & \nd & 0.91(0.15) & 
        1.09(0.20) & \nd & \nd \nl
\enddata

\tablenotetext{a}{From positions cited in \cite{kbs85} and \cite{wills97}}
\tablenotetext{b}{Only clearly isolated sources were measured}

\end{deluxetable}

\clearpage

%% TABLE 3

\begin{deluxetable}{rr}
%\small
\tablewidth{0pt}
\tablecaption{Estimated Fluxes for 41.9+58 (Epoch 1993.4) \label{tbl-3}}
\tablehead{\colhead{Frequency} &
\colhead{Flux} \nl
\colhead{(GHz)} &
\colhead{(mJy)}
}
\startdata
 0.4080         & 117.10 (7.6)  \nl
 1.4524         &  88.46 (2.4)  \nl
 1.666\phm{0}   &  79.00 (4.0)  \nl
 4.8351         &  40.02 (0.9)  \nl
 4.995\phm{0}   &  34.00 (4.0)  \nl
 8.4141         &  26.76 (0.5)  \nl
14.9649         &  17.88 (1.7)  \nl
22.4601         &  12.01 (0.6)  \nl
\enddata
\end{deluxetable}

\clearpage

%%TABLE 4

\begin{deluxetable}{cccccccc}
\footnotesize
\tablewidth{0pt}
\tablecaption{Properties of Radio Sources - No Absorption \label{tbl-4}}
\tablehead{\colhead{Source name} & 
\colhead{I$_{sy}$} & 
\colhead{I$_{th}$} &
\colhead{$\alpha$} & 
\colhead{$\nu_{\tau=1}$} &
\colhead{Diameter\tablenotemark{a}} &
\colhead{B-field\tablenotemark{b}} &
\colhead{Pressure\tablenotemark{c}} \nl 
 & \colhead{(mJy)} &
\colhead{(mJy)} & &
\colhead{(GHz)} &
\colhead{(arcsec)} &
\colhead{(mG)} &
\colhead{(dyne cm$^{-2}$)}
}
\startdata
39.11+573 &  8.62 & 0.0     & -0.38 (0.01) & 0.0
        & 0.140 & 2.35 & 17.1\phn \nl
39.40+561 &  3.05 & 0.0     & -0.21 (0.01) & 0.0 
        & 0.200 & 1.37 & \phn5.85 \nl
39.77+569 &  2.13 & 0.0     & -0.49 (0.01) & 0.0 
        & 0.200 & 1.14 & \phn4.06 \nl
40.32+551 &  2.29 & 0.0     & -0.55 (0.01) & 0.0 
        & 0.200 & 1.17 & \phn4.25 \nl
44.28+592 &  6.65 & 0.0     & -0.56 (0.01) & 0.0
        & 0.140 & 2.16 & 14.4\phn \nl
45.25+651 &  5.42 & 0.0     & -0.62 (0.01) & 0.0
        & 0.220 & 1.40 & \phn6.05 \nl
45.43+673 &  4.69 & 0.0     & -0.57 (0.01) & 0.0 
        & 0.145 & 1.90 & 11.2\phn \nl
45.48+648 &  1.67 & 0.0     & -0.15 (0.01) & 0.0 
        & 0.200 & 1.20 & \phn4.44 \nl
45.74+652 &  5.14 & 0.0     & -0.23 (0.01) & 0.0
        & 0.200 & 1.58 & \phn7.72 \nl
47.37+680 &  3.16 & 0.0     & -0.57 (0.01) & 0.0  
        & 0.200 & 1.29 & \phn5.13 \nl
\enddata

\tablenotetext{a}{From \cite{muxlow94} where available, otherwise
values are set to 0.200 arcsec}
\tablenotetext{b}{$\times$ (1+k)$^{2/7}$ $\phi^{-2/7}$}
\tablenotetext{c}{$\times$10$^{-8}$ (1+k)$^{4/7}$ $\phi^{3/7}$ }

\end{deluxetable}

\clearpage

%% TABLE 5

\begin{deluxetable}{crlrrcrr}
\footnotesize
\tablewidth{0pt}
\tablecaption{Properties of Radio Sources - Screen Model Absorption 
\label{tbl-5}}
\tablehead{\colhead{Source name} & 
\colhead{I$_{sy}$} & 
\colhead{I$_{th}$} &
\colhead{$\alpha$} & 
\colhead{$\nu_{\tau=1}$} &
\colhead{Diameter\tablenotemark{a}} &
\colhead{B-field\tablenotemark{b}} &
\colhead{Pressure\tablenotemark{c}} \nl 
 & \colhead{(mJy)} &
\colhead{(mJy)} & &
\colhead{(GHz)} &
\colhead{(arcsec)} &
\colhead{(mG)} &
\colhead{(dyne cm$^{-2}$)}
}
\startdata
39.64+534 &  4.42 & 1.34e-3 & -0.71 (0.01) & 0.45 (0.01) 
        & 0.200 & 1.47 & 6.72 \nl
40.68+550 & 17.89 & 5.04e-4 & -0.52 (0.01) & 0.40 (0.01) 
        & 0.240 & 1.80 & 10.0\phn \nl
41.31+596 &  8.59 & 4.49e-2 & -0.54 (0.01) & 1.11 (0.01)
        & 0.140 & 2.32 & 16.6\phn \nl
41.96+574 &122.76 & 1.24e-4 & -0.72 (0.01) & 0.34 (0.01)
        & 0.021 & 26.4\phn & 2160.\phn\phn \nl
42.21+590 & 0.00  & 5.20    & -0.10 (0.01) & 2.40 (0.01)
        & 0.275 & \nd & \nd \nl
42.53+619 & 30.92 & 2.51e-2 & -1.84 (0.01) & 0.65 (0.01) 
        & 0.110 & 11.7\phn & 422.\phn\phn \nl
42.67+556 &  4.44 & 1.12e-3 & -0.61 (0.01) & 0.42 (0.01) 
        & 0.195 & 1.46 & 6.61 \nl
43.19+583 & 15.31 & 8.68e-4 & -0.67 (0.01) & 0.45 (0.01)
        & 0.075 & 4.79 & 71.3\phn \nl
43.31+591 & 30.31 & 8.89e-5 & -0.64 (0.01) & 0.24 (0.01)
        & 0.195 & 2.54 & 20.1\phn \nl
44.01+595 & 61.99 & 2.66e-1 & -0.51 (0.01) & 2.01 (0.01)
        & 0.050 & 9.83 & 300.\phn\phn \nl
44.52+581 &  7.21 & 3.83e-3 & -0.61 (0.01) & 0.56 (0.01)
        & 0.240 & 1.40 & 6.06 \nl
45.18+612 & 24.08 & 4.19e-4 & -0.68 (0.01) & 0.38 (0.01)
        & 0.225 & 2.13 & 14.1\phn \nl
45.86+640 &  4.10 & 6.52e-4 & -0.53 (0.01) & 0.34 (0.01) 
        & 0.070 & 3.39 & 35.7\phn \nl
46.52+638 &  9.71 & 9.74e-4 & -0.73 (0.01) & 0.46 (0.01)
        & 0.250 & 1.53 & 7.30 \nl
46.56+738 &  3.66 & 4.57e-4 & -0.78 (0.01) & 0.34 (0.01) 
        & 0.200 & 1.44 & 6.43 \nl
46.70+670 &  5.22 & 4.15e-4 & -0.57 (0.01) & 0.36 (0.01) 
        & 0.220 & 1.37 & 5.81 \nl
47.37+680 &  4.14 & 4.58e-3 & -0.83 (0.01) & 0.37 (0.01) 
        & 0.200 & 1.53 & 7.28 \nl
\enddata

\tablenotetext{a}{From \cite{muxlow94} where available, otherwise
values are set to 200 mas}
\tablenotetext{b}{$\times$ (1+k)$^{2/7}$ $\phi^{-2/7}$}
\tablenotetext{c}{$\times$10$^{-8}$ (1+k)$^{4/7}$ $\phi^{3/7}$ }

\end{deluxetable}

\clearpage

%% TABLE 6

\begin{deluxetable}{crcrrrcrr}
%%\small
%%\footnotesize
\scriptsize
\tablewidth{0pt}
\tablecaption{Values of Fit Parameters for Radio Sources - Mixed 
Absorption Model \label{tbl-6}}
\tablehead{\colhead{Source name} & 
\colhead{I$_{sy}$} & 
\colhead{Flux$_{thin}$} &
\colhead{I$_{th}$} &
\colhead{$\alpha$} & 
\colhead{$\nu_{\tau=1}$} &
\colhead{Diameter\tablenotemark{b}} &
\colhead{B-field\tablenotemark{c}} &
\colhead{Pressure\tablenotemark{d}} \nl 
 & \colhead{(mJy)} &
\colhead{at 1 GHz (mJy)\tablenotemark{a}} &
\colhead{(mJy)} & &
\colhead{(GHz)} &
\colhead{(arcsec)} &
\colhead{(mG)} &
\colhead{(dyne cm$^{-2}$)}
}
\startdata
40.68+550 &  44.08 & \phn19.01  & 2.33e-5 
        & -0.55 (0.01) & 0.67 (0.01) & 0.240 & 1.83 & 10.4\phn \nl
41.31+596 &   3.05 &  \phn\phn9.07  & 3.35e-3
        & -0.56 (0.01) & 1.68 (0.01) & 0.140 & 2.36 & 17.2\phn \nl
41.96+574 & 463.35 & 127.04 & 1.61e-1 
        & -0.74 (0.01) & 0.54 (0.01) & 0.021 & 26.8\phn & 2240.\phn\phn \nl
42.53+619 &  16.03 & \phn42.45  & 2.44e-3
        & -1.99 (0.01) & 1.59 (0.01) & 0.110 & 15.1\phn & 706.\phn\phn\nl
42.67+556 &  10.66 &  \phn\phn4.60  & 8.02e-3 
        & -0.63 (0.01) & 0.67 (0.01) & 0.195 & 1.48 & 6.81 \nl
43.19+583 &  30.17 & \phn15.58  & 3.95e-1 
        & -0.72 (0.01) & 0.73 (0.01) & 0.075 & 4.91 & 74.7\phn \nl
43.31+591 & 156.73 & \phn29.30  & 3.94 
        & -0.85 (0.01) & 0.45 (0.01) & 0.195 & 2.77 & 23.8\phn \nl
44.01+595 &   4.48 & \phn72.71  & 5.32e-2 
        & -0.56 (0.01) & 3.77 (0.01) & 0.050 & 10.3\phn & 331.\phn\phn \nl
44.52+581 &   8.63 &  \phn\phn7.58  & 2.10e-3 
        & -0.64 (0.01) & 0.94 (0.01) & 0.240 & 1.43 & 6.38 \nl
45.18+612 &  77.12 & \phn24.57  & 2.37e-3 
        & -0.69 (0.01) & 0.58 (0.01) & 0.225 & 2.15 & 14.4\phn \nl
45.86+640 &  18.60 &  \phn\phn4.16  & 3.36e-3
        & -0.54 (0.01) & 0.49 (0.01) & 0.070 & 3.41 & 36.1\phn \nl
46.52+638 &   9.07 & \phn13.53  & 1.84  
        & -1.50 (0.01) & 1.21 (0.01) & 0.250 & 3.17 & 31.1\phn \nl
46.56+738 &  10.08 &  \phn\phn3.69  & 3.93e-1
        & -1.04 (0.01) & 0.62 (0.01) & 0.200 & 1.72 & 9.16 \nl
46.70+670 &  11.33 &  \phn\phn4.89  & 1.08 
        & -0.90 (0.01) & 0.67 (0.01) & 0.220 & 1.55 & 7.43 \nl
\enddata

\tablenotetext{a}{Flux = $I_{sy} \nu^{(2.1-\alpha)}\nu_{\tau=1}^{2.1}$}
\tablenotetext{b}{\cite{muxlow94}}
\tablenotetext{c}{$\times$ (1+k)$^{2/7}$ $\phi^{-2/7}$}
\tablenotetext{d}{$\times$10$^{-8}$ (1+k)$^{4/7}$ $\phi^{3/7}$}

\end{deluxetable}

\clearpage

%% TABLE 7

\begin{deluxetable}{ll}
\tablewidth{0pt}
\tablecaption{Derived Properties of a Discrete \HII Region \label{tbl-7}}
\tablehead{}
\startdata
Source Name                                & 42.21+590     \nl
Diameter\tablenotemark{a}                  & 0.275 \arcsec \nl
Optically thin flux at 10 GHz (model fit)  & 4.398 mJy     \nl
Optical Depth at 1 GHz (model fit)         & 6.287         \nl
Optically thick flux at 1 GHz (model fit)  & 0.901 mJy     \nl
Frequency where $\tau$ = 1 (model fit)     & 2.40 GHz      \nl
Brightness Temperature (1 GHz)             & 2.10$\times$10$^4$ K  \nl
%%Brightness Temperature (10 GHz)            & 1.025$\times$10$^3$ K \nl
%%Electron Temperature (1 GHz)               & 2.10$\times$10$^4$ K  \nl
%%Electron temperature (10 GHz)              & 2.10$\times$10$^4$ K  \nl
Thermal Luminosity                   & 6.932 $\times$10$^{18}$ W Hz$^{-1}$\nl
Production Rate of Lyman continuum photons & 2.43$\times$10$^{50}$ s$^{-1}$ \nl
Emission Measure at 1 GHz                  & 5.00$\times$10$^7$ cm$^{-6}$ pc\nl
Equivalent Number of O8 Stars              & 108 \nl
%%Predicted H$\beta$ Flux (unextinguished)   & 
%%      5.07$\times$10$^{-12}$ erg cm$^{-2}$ s$^{-1}$ \nl
\enddata
\tablenotetext{a}{From \cite{muxlow94}}
\end{deluxetable}

%% Figures and figure captions

\clearpage

\figcaption[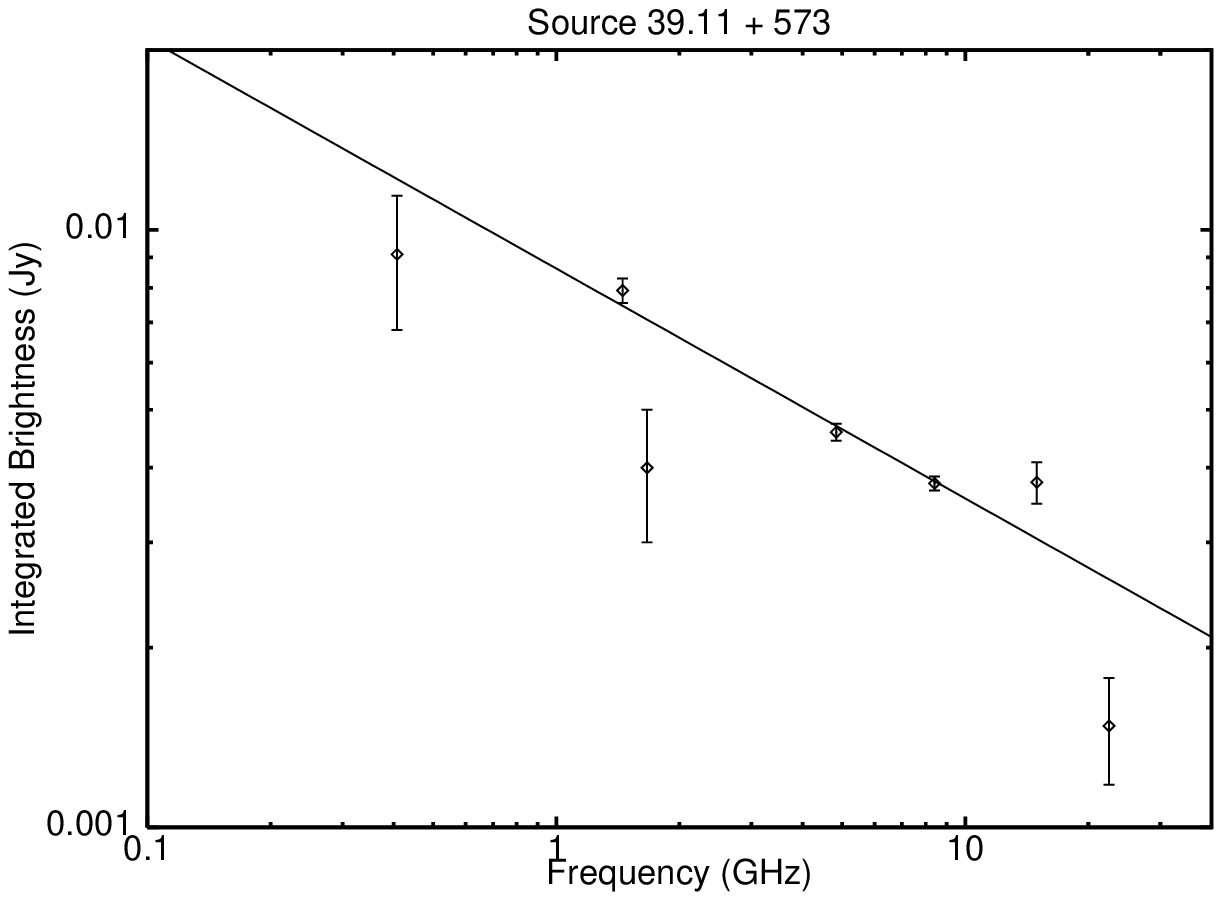]{Radio spectra of individual point sources.  For 
figures with two lines, the solid line represents the best fit model of
an intervening screen of thermal absorption between the source and the 
observer; the dotted line represents a model where the absorbing and emitting 
gas is mixed.  Source 47.37+680 is the exception (see text).
Upper flux limits are indicated by unlabelled arrows.}

\figcaption[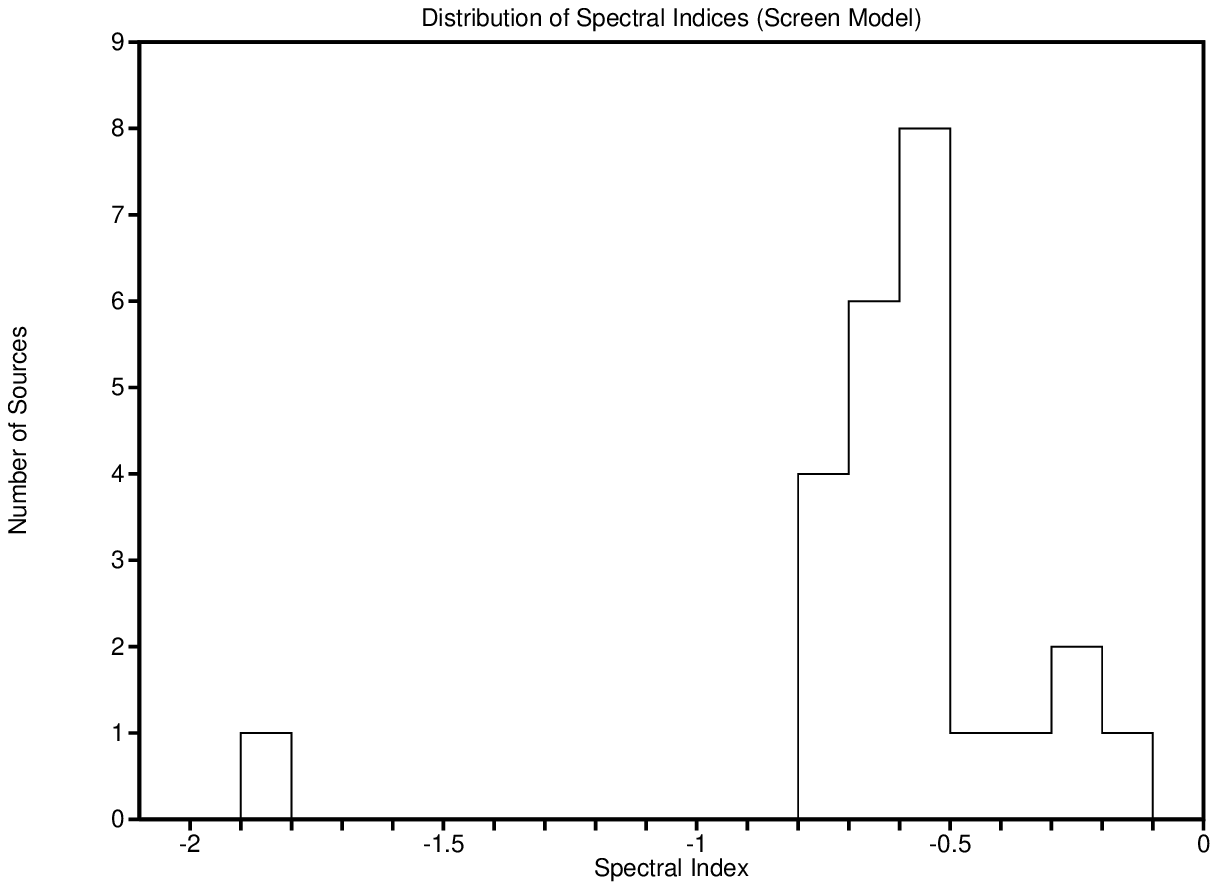]{Histogram indicating the distribution of the 
non-thermal spectral indices of the 24 point sources for which full spectra
were obtained, using the screen-absorption model.  Fits where no absorption 
was found are included.}

\figcaption[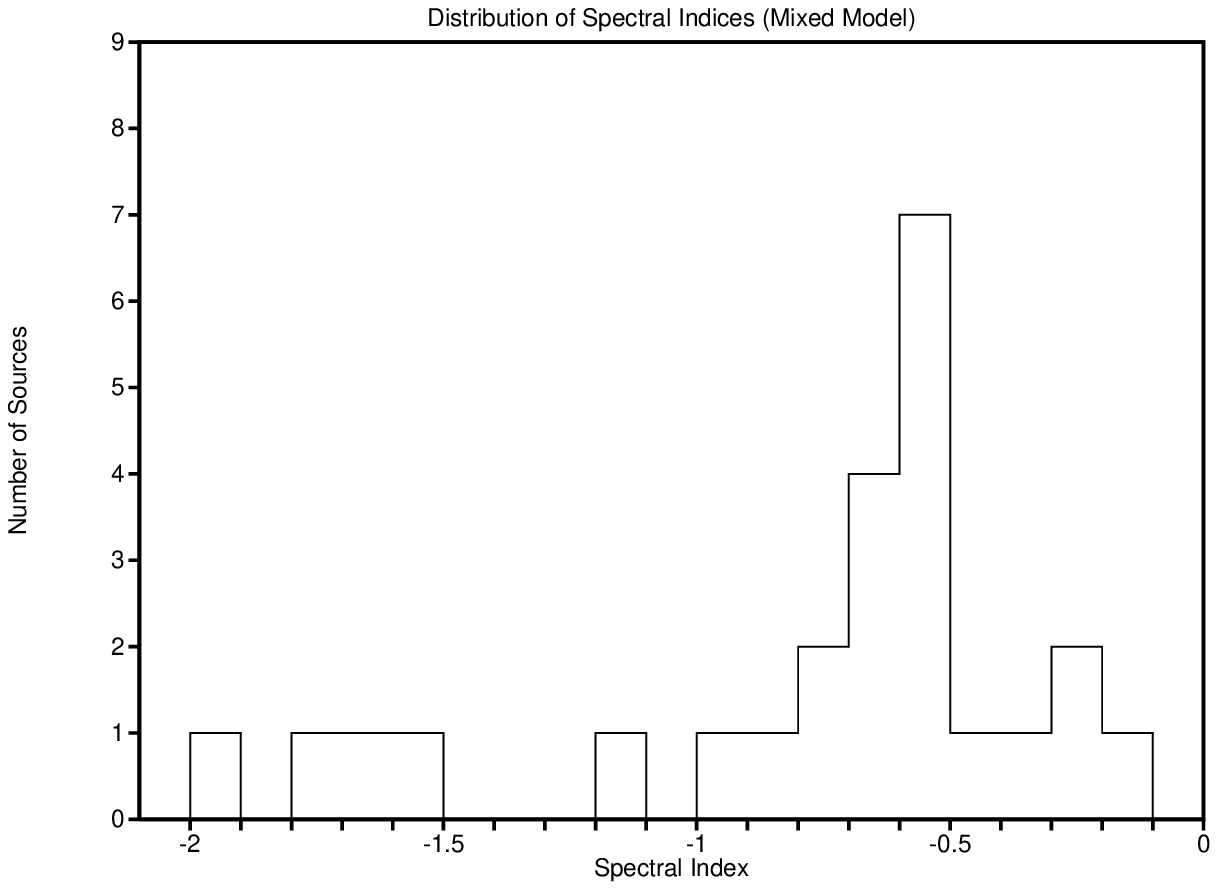]{Histogram indicating the distribution of the
non-thermal spectral indices of the 23 point sources where full spectra were
obtained, using the co-spatial absorption model. Fits where no absorption
was found are included.} 

\figcaption[SNR47.4+680.PS]{Shell-like source 47.37+680, from a 3.6 cm image.
Contours are 9, 10, 11 ... 22 $\times$ 60 $\mu$Jy arcsec$^{-2}$. Grey scale
is 100-220 $\mu$Jy beam$^{-1}$.  The beam is shown at lower left.}

\clearpage

%5FIGURE 1

\begin{figure}
\figurenum{}
\plotfiddle{snr1_fn.eps}{0pt}{0}{70}{70}{-300}{-46}
\plotfiddle{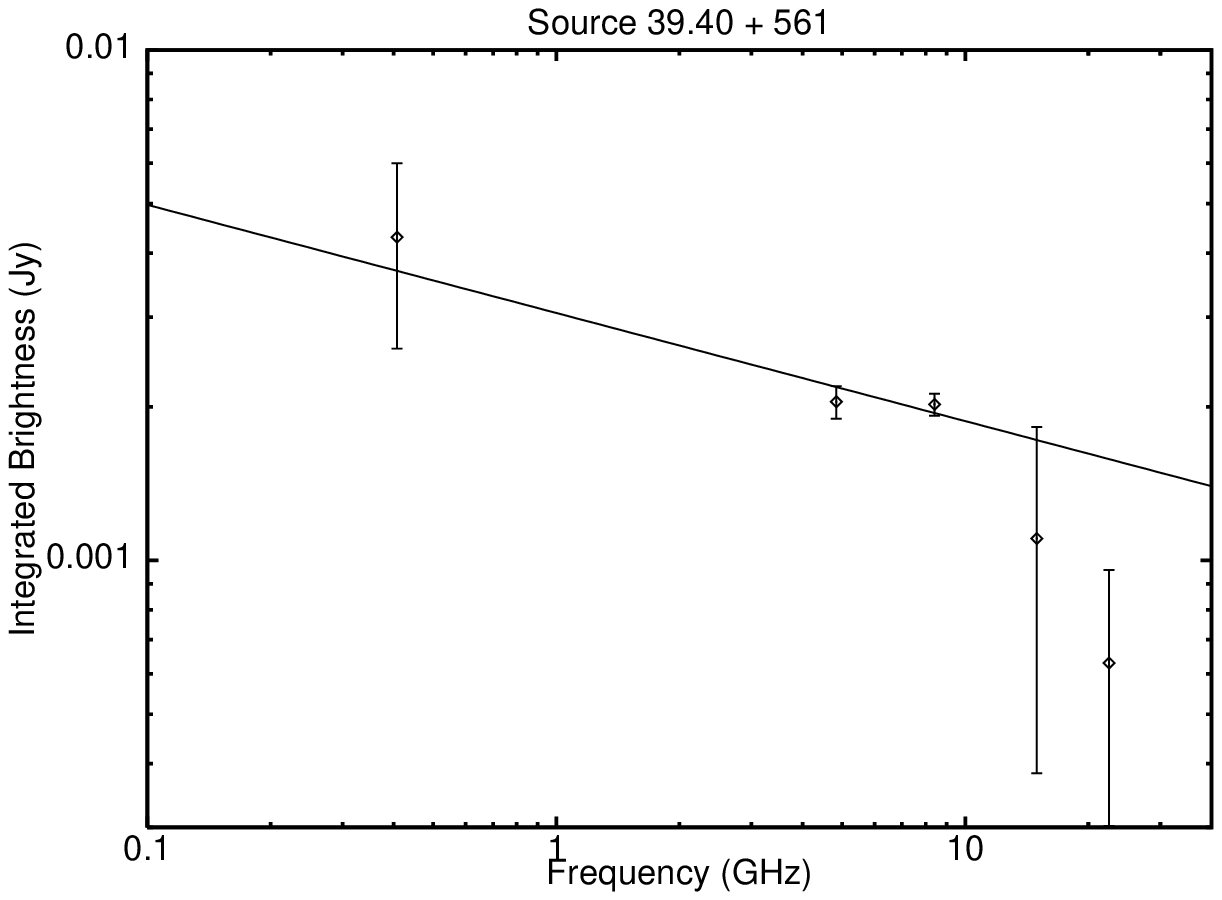}{0pt}{0}{70}{70}{-20}{0}
\plotfiddle{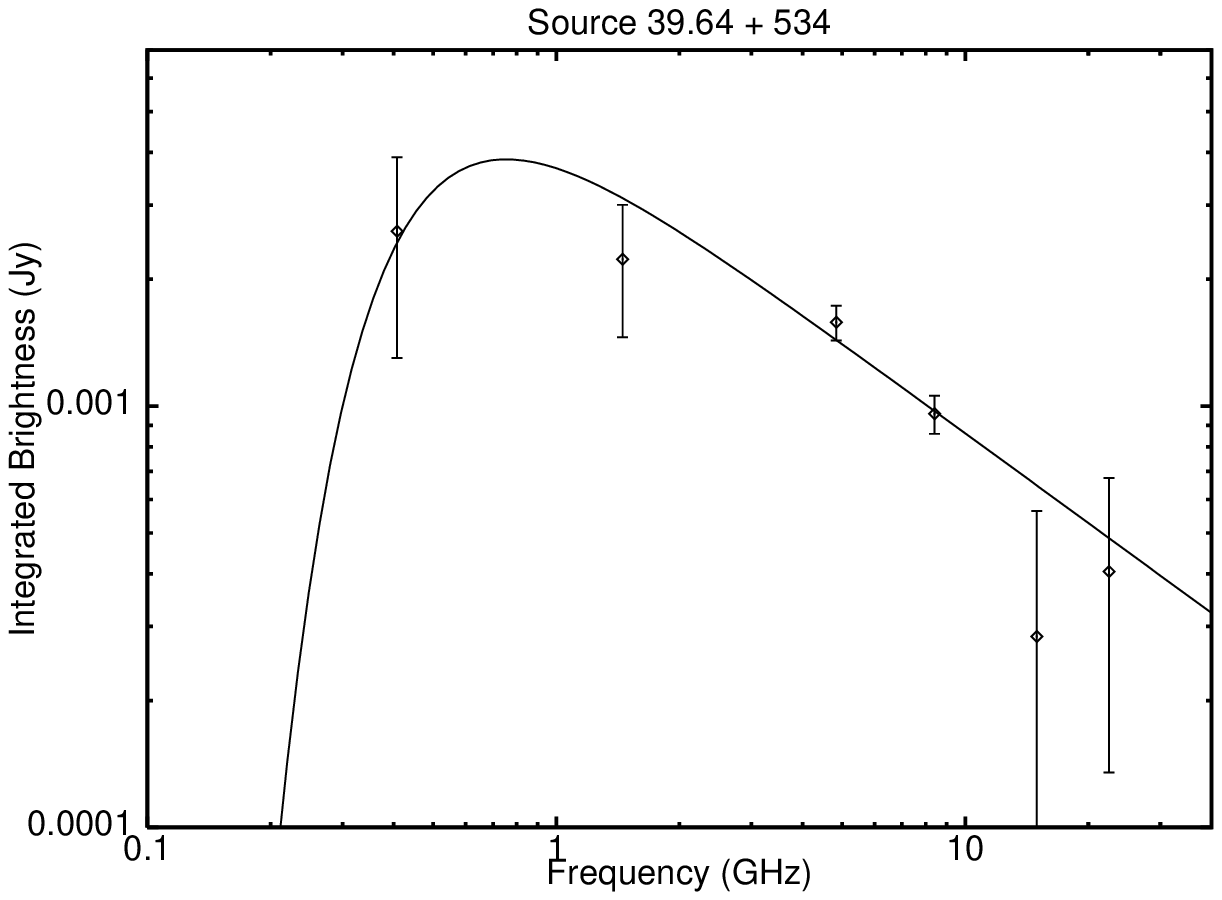}{0pt}{0}{70}{70}{-300}{-141}
\plotfiddle{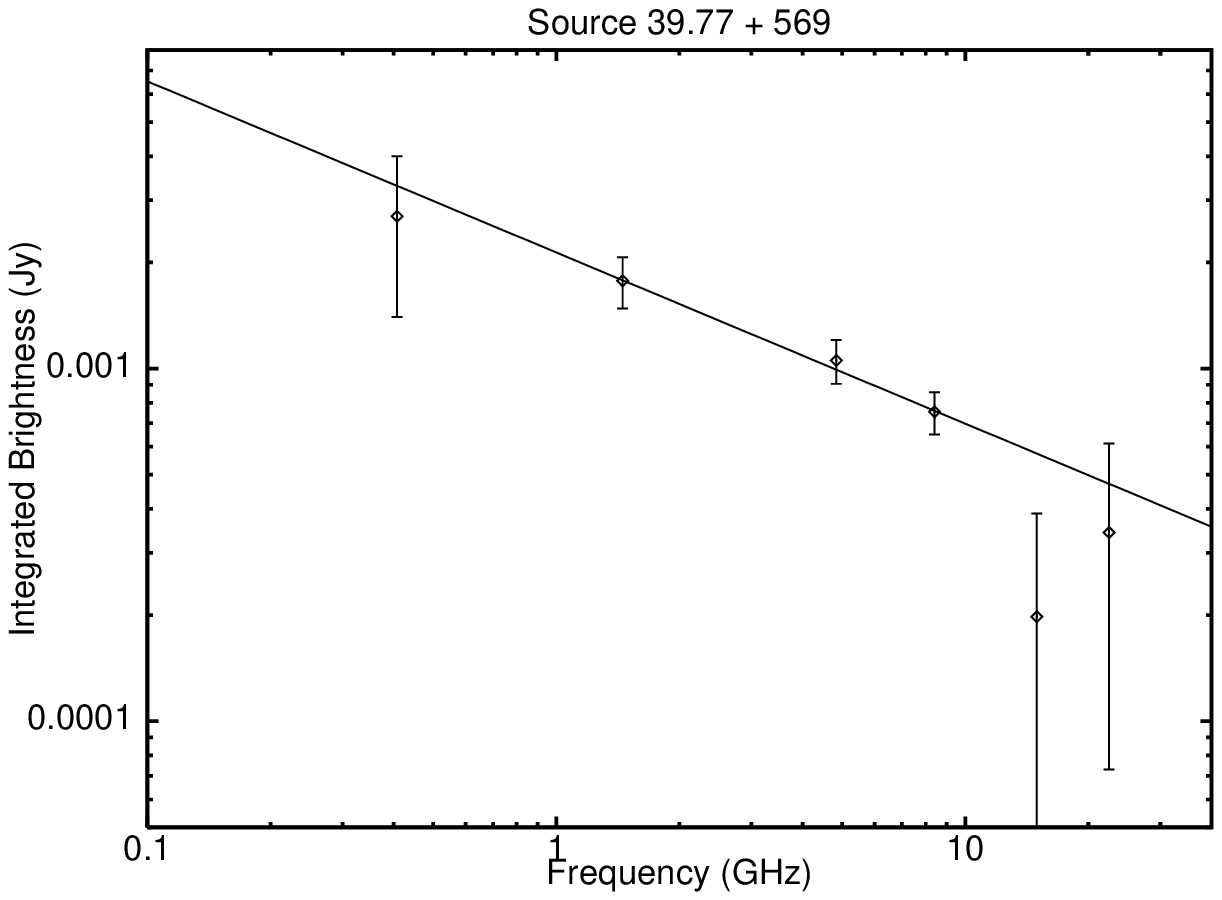}{0pt}{0}{70}{70}{-20}{-95}
\plotfiddle{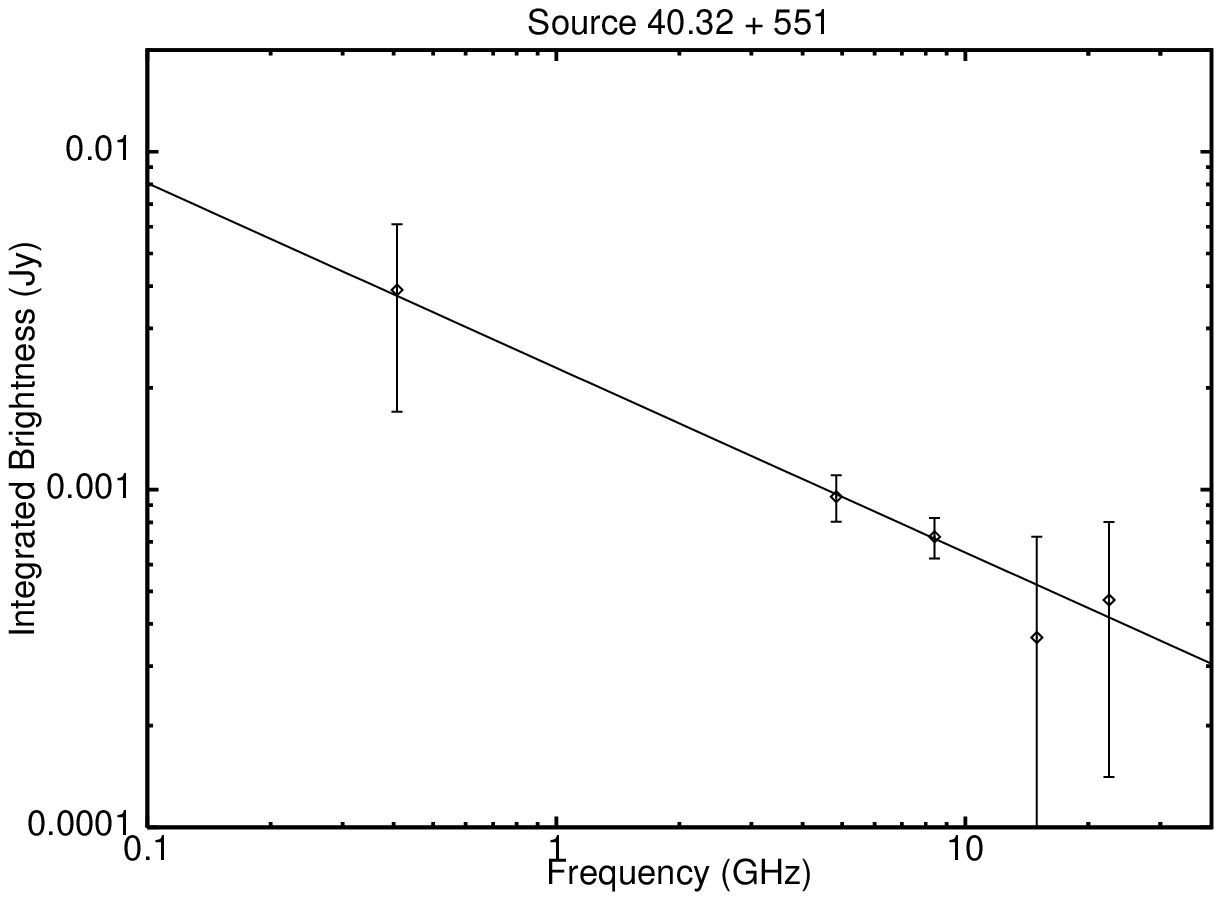}{0pt}{0}{70}{70}{-300}{-236}
\plotfiddle{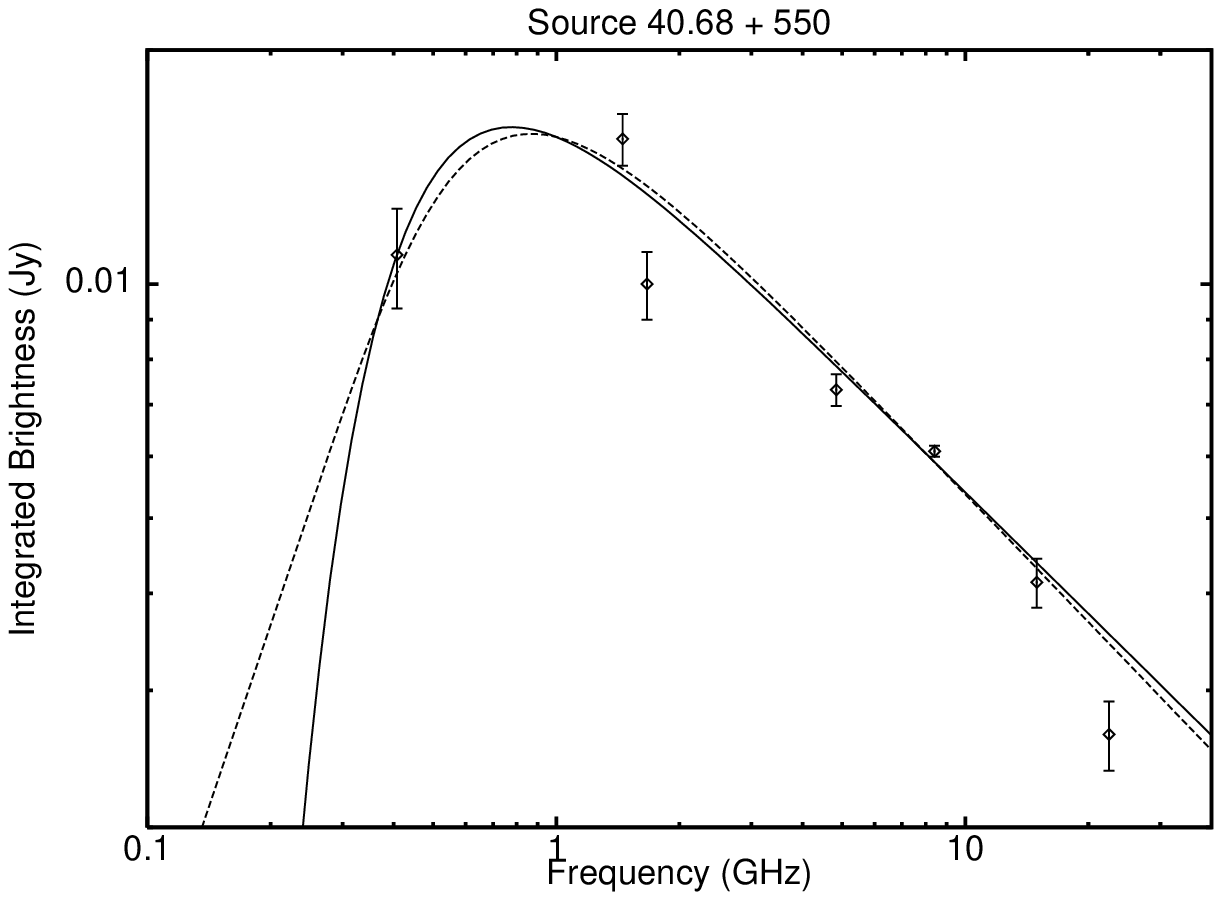}{0pt}{0}{70}{70}{-20}{-190}
\end{figure}

\clearpage

\begin{figure}
\figurenum{}
\plotfiddle{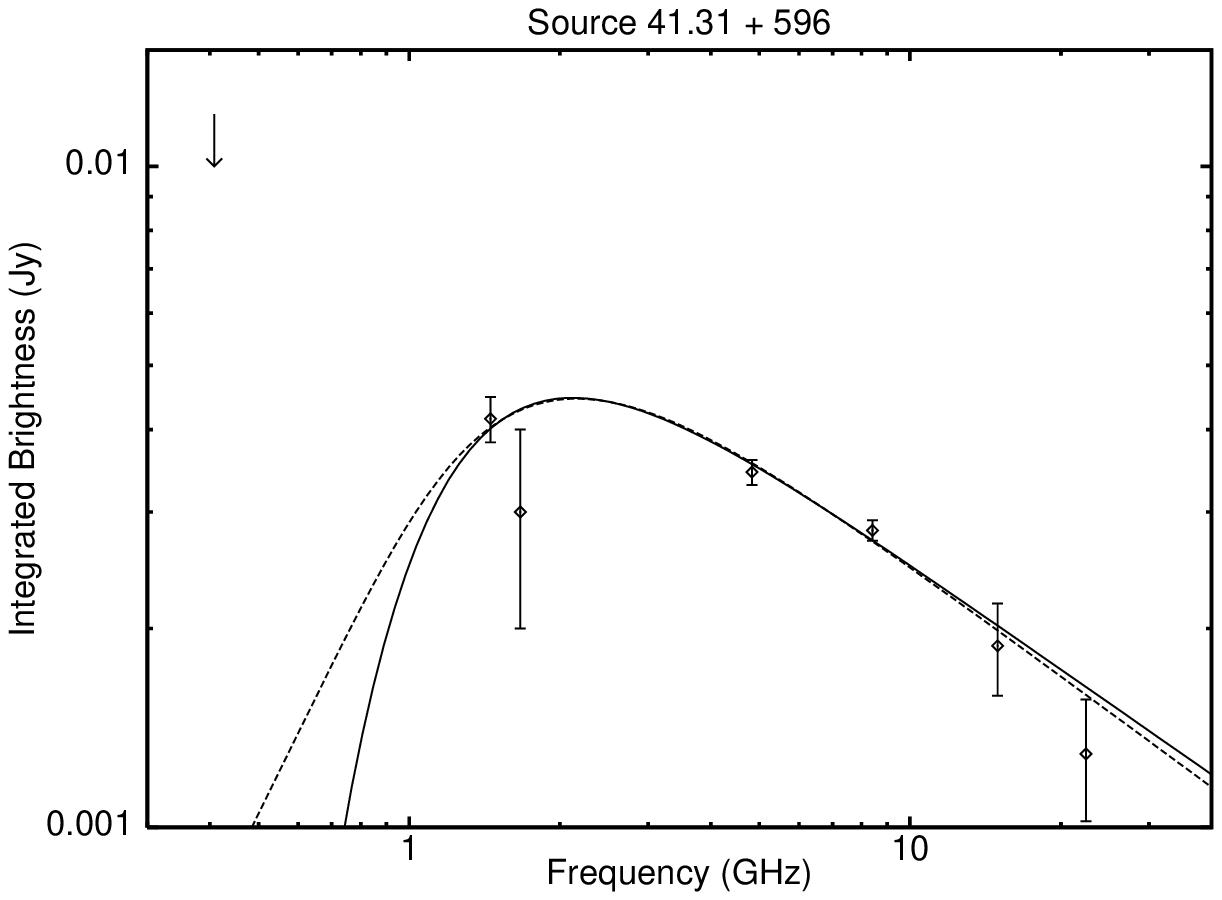}{0pt}{0}{70}{70}{-300}{-46}
\plotfiddle{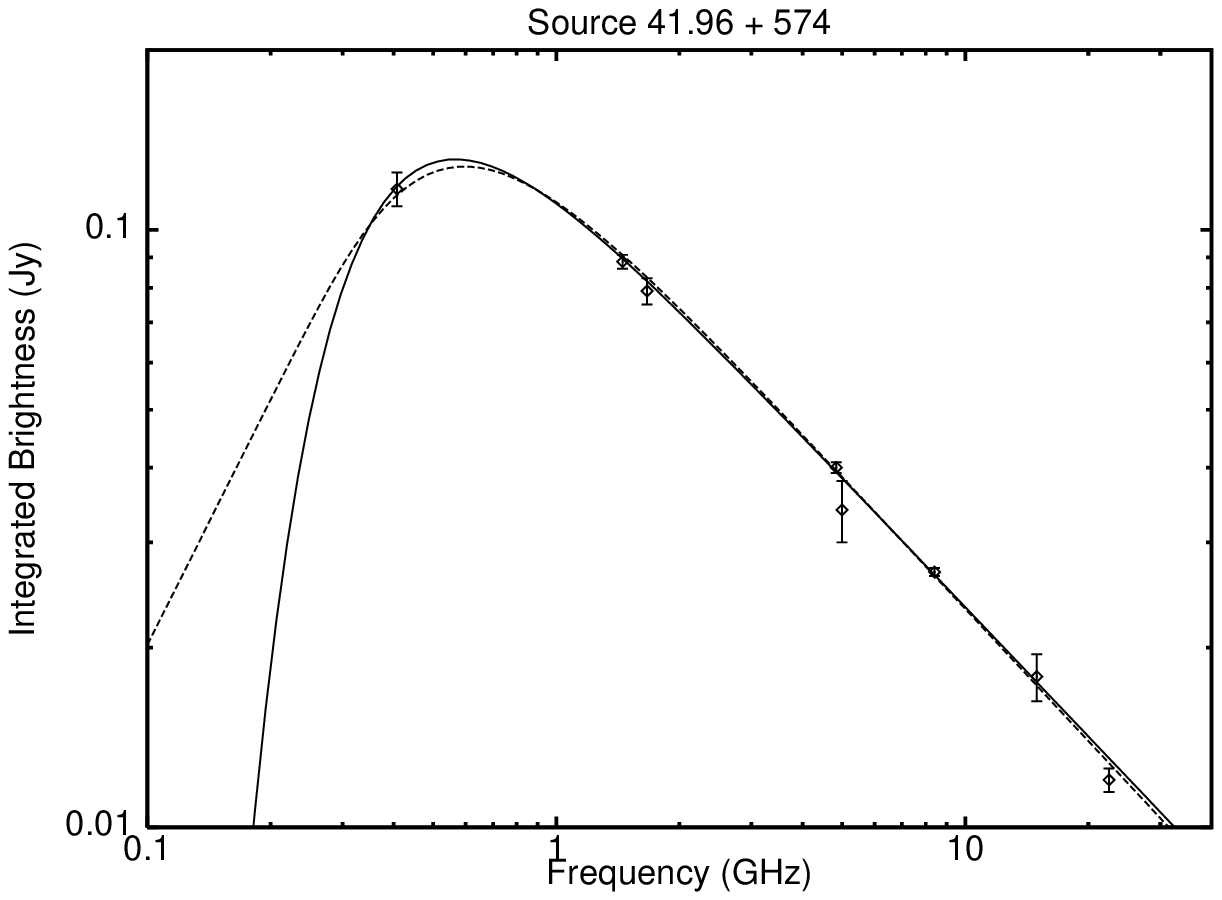}{0pt}{0}{70}{70}{-20}{0}
\plotfiddle{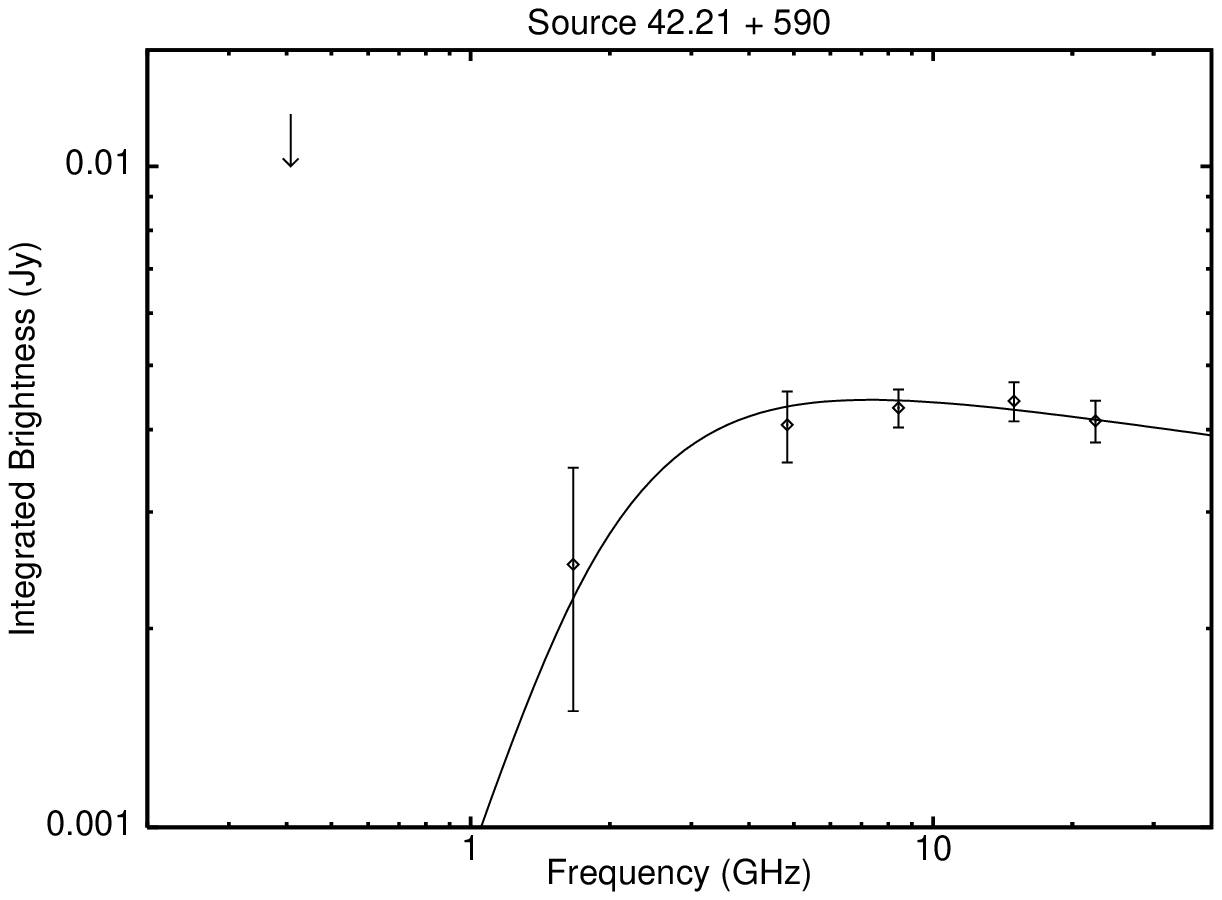}{0pt}{0}{70}{70}{-300}{-141}
\plotfiddle{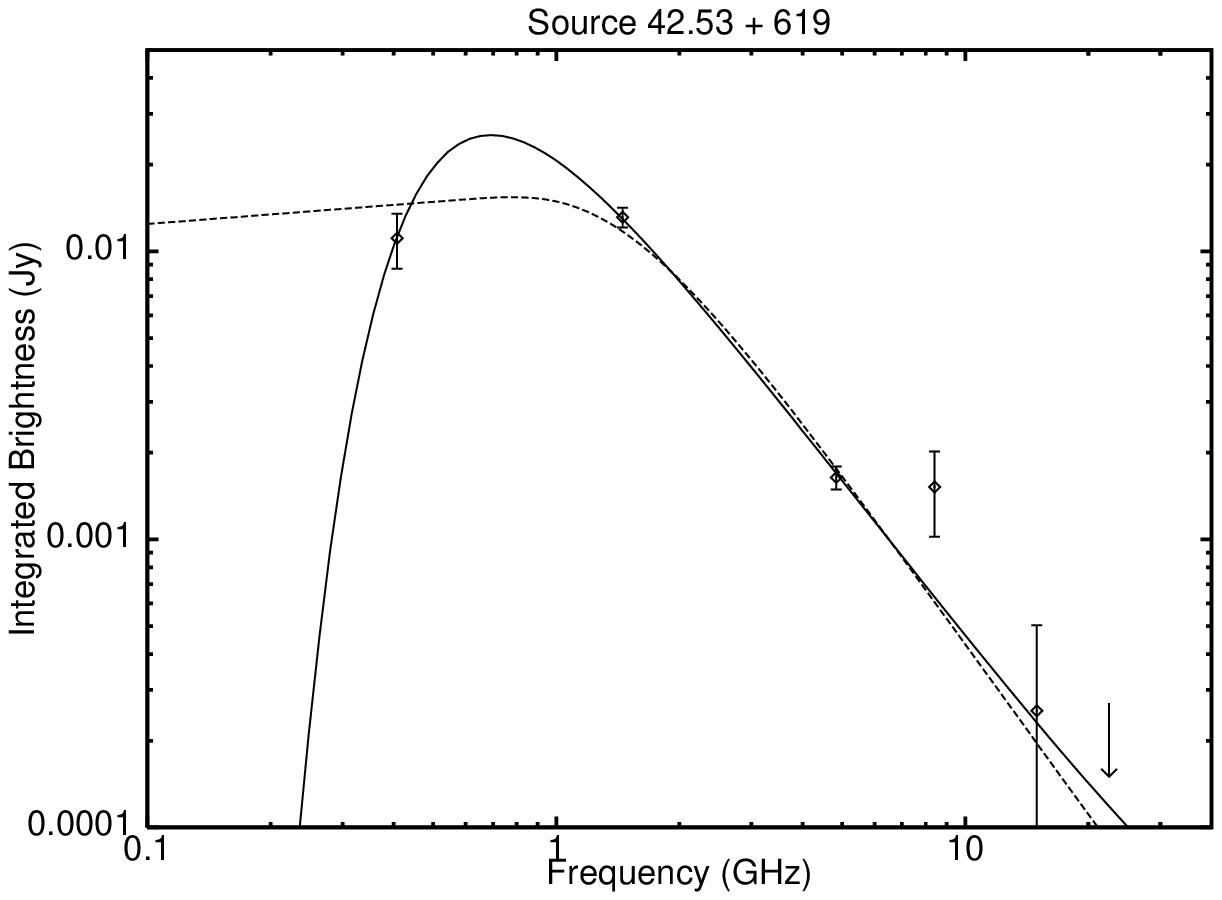}{0pt}{0}{70}{70}{-20}{-95}
\plotfiddle{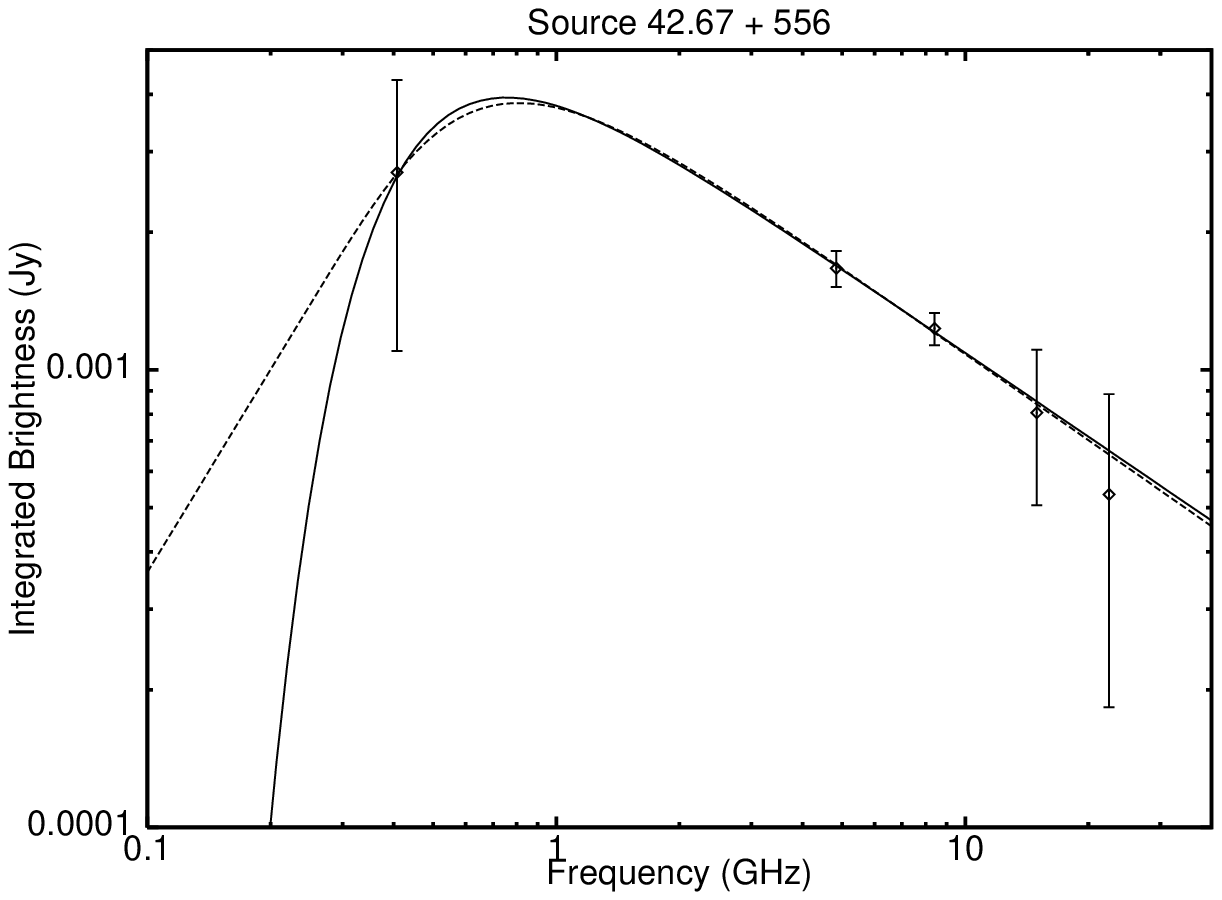}{0pt}{0}{70}{70}{-300}{-236}
\plotfiddle{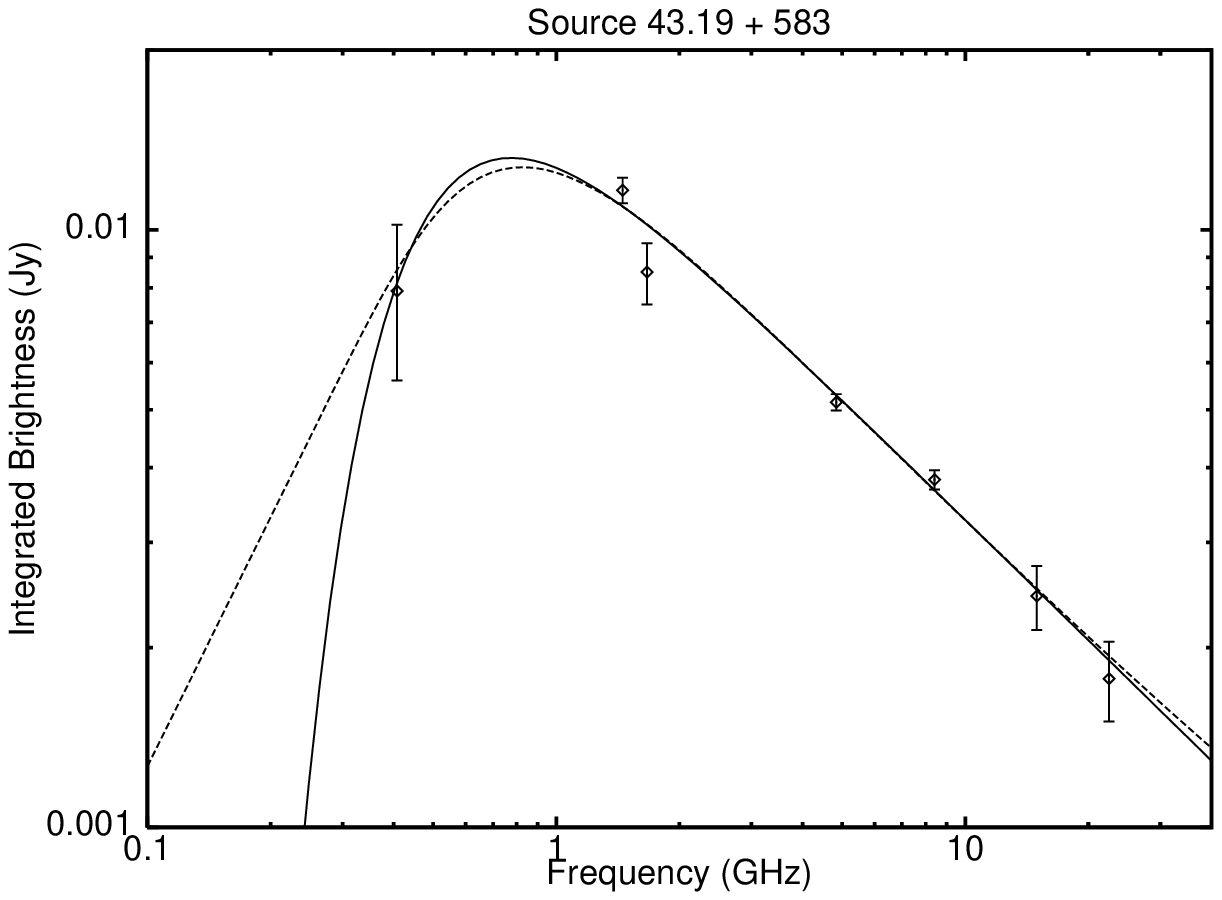}{0pt}{0}{70}{70}{-20}{-190}
\end{figure}

\clearpage

\begin{figure}
\figurenum{}
\epsscale{1.0}
\plotfiddle{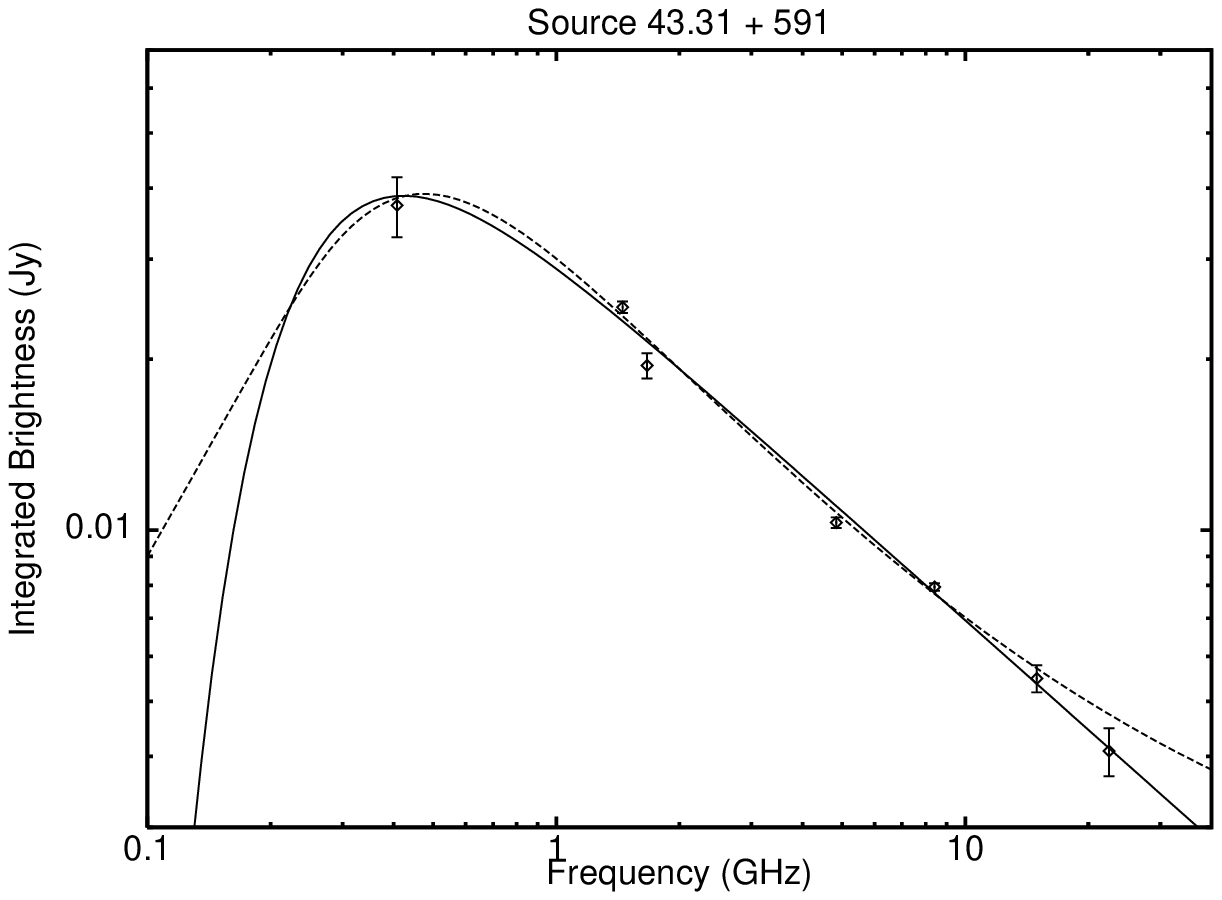}{0pt}{0}{70}{70}{-300}{-46}
\plotfiddle{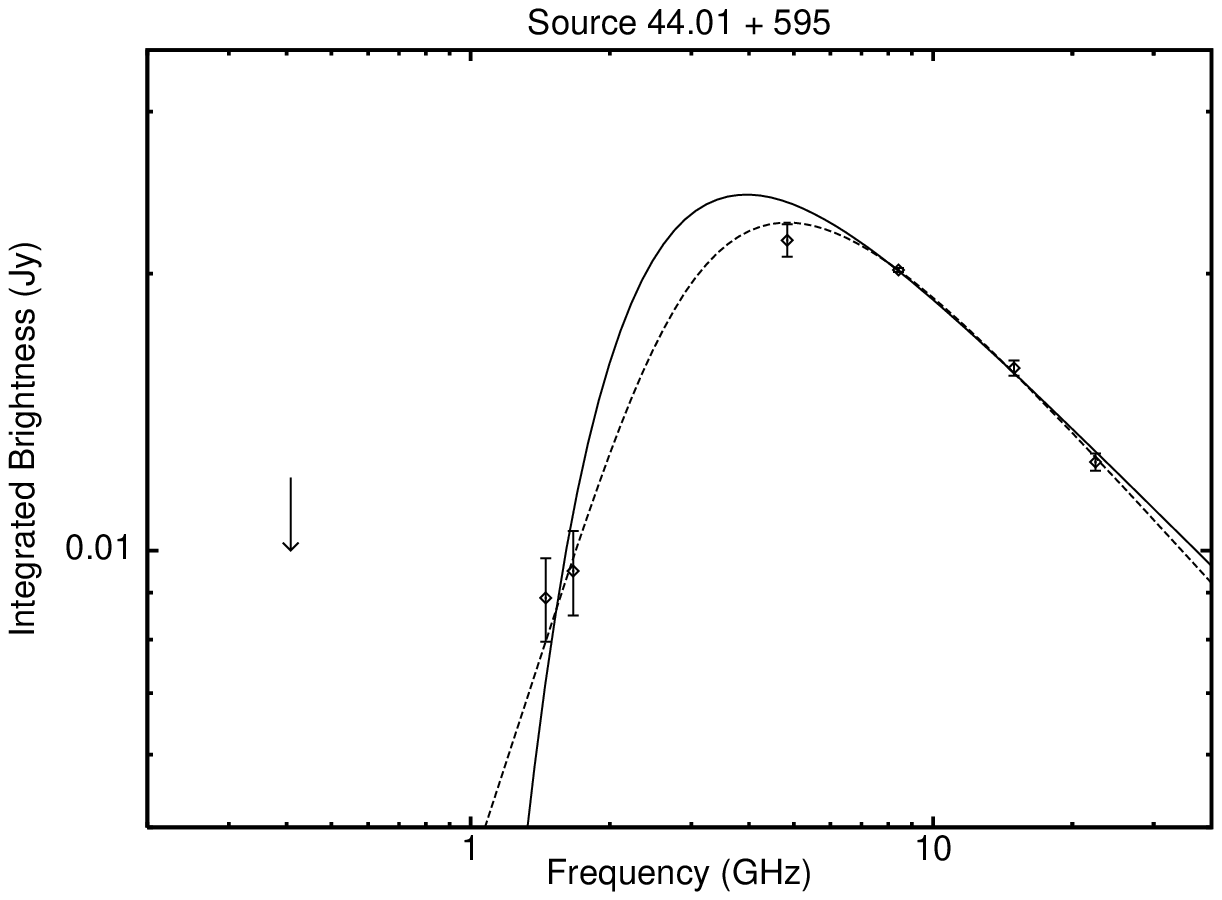}{0pt}{0}{70}{70}{-20}{0}
\plotfiddle{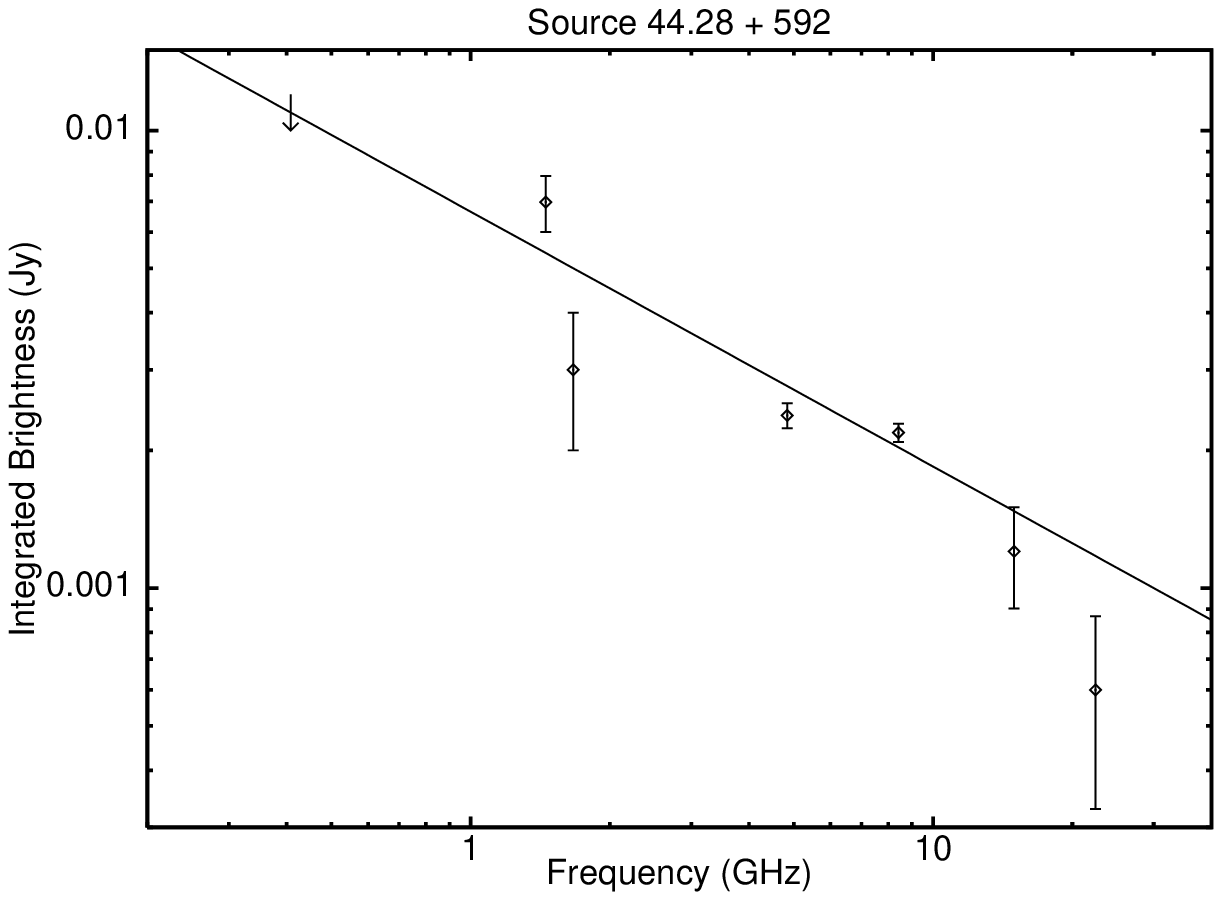}{0pt}{0}{70}{70}{-300}{-141}
\plotfiddle{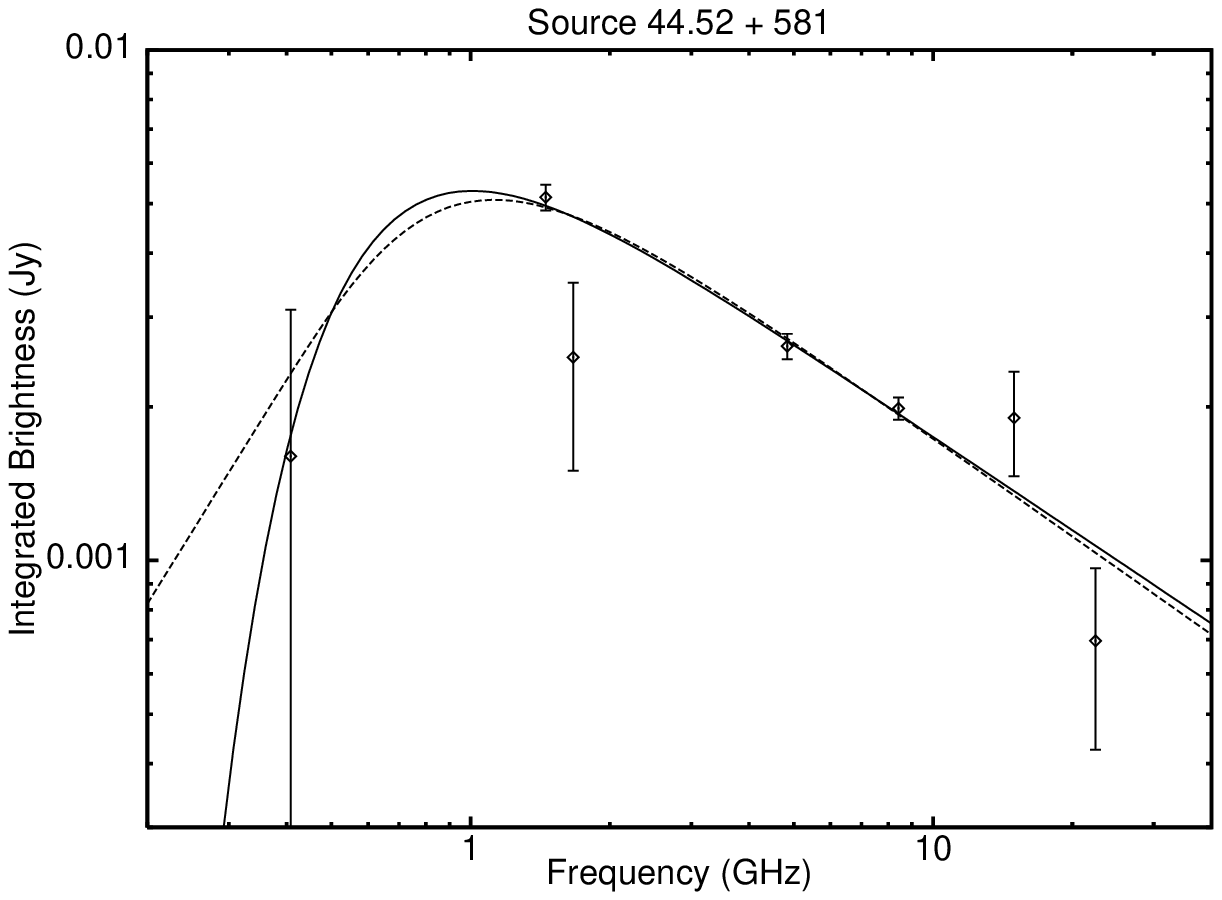}{0pt}{0}{70}{70}{-20}{-95}
\plotfiddle{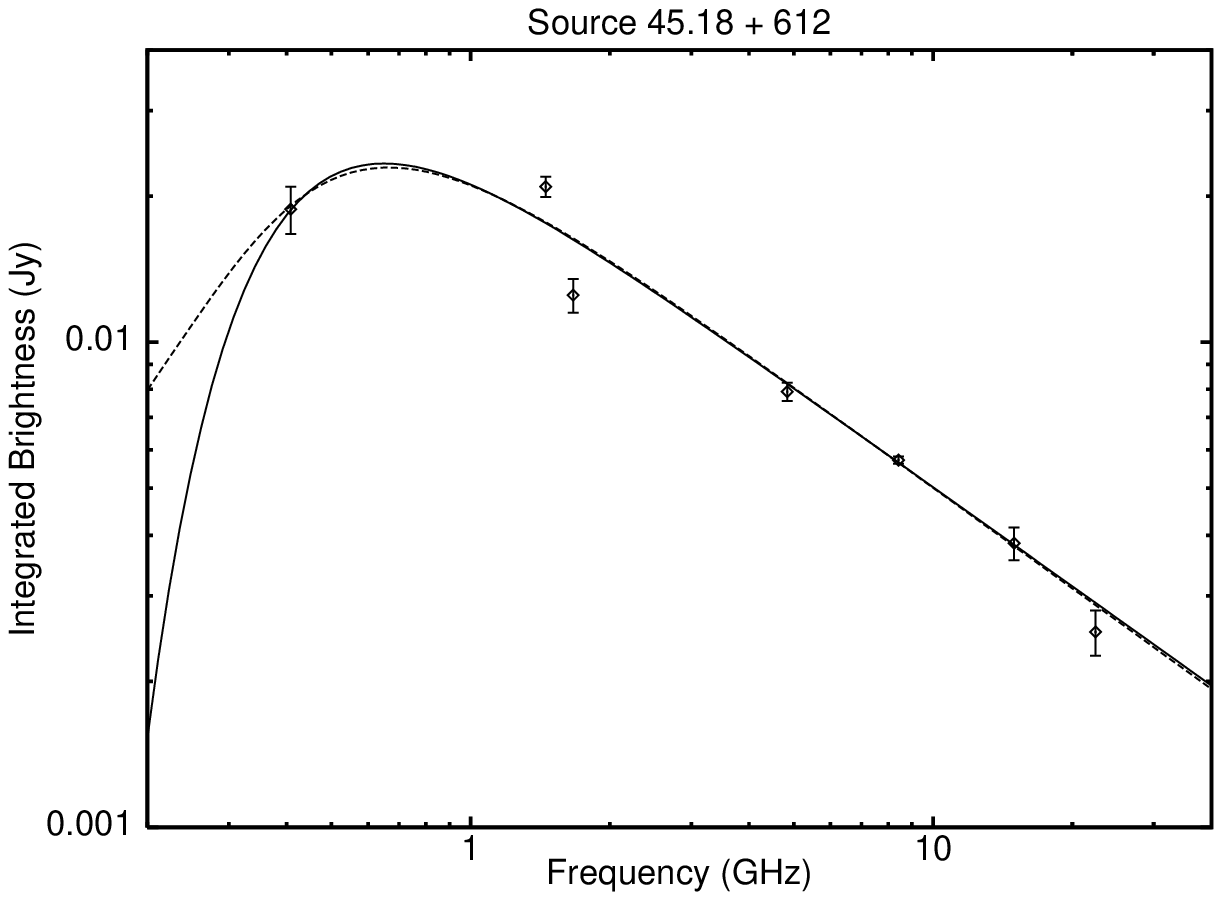}{0pt}{0}{70}{70}{-300}{-236}
\plotfiddle{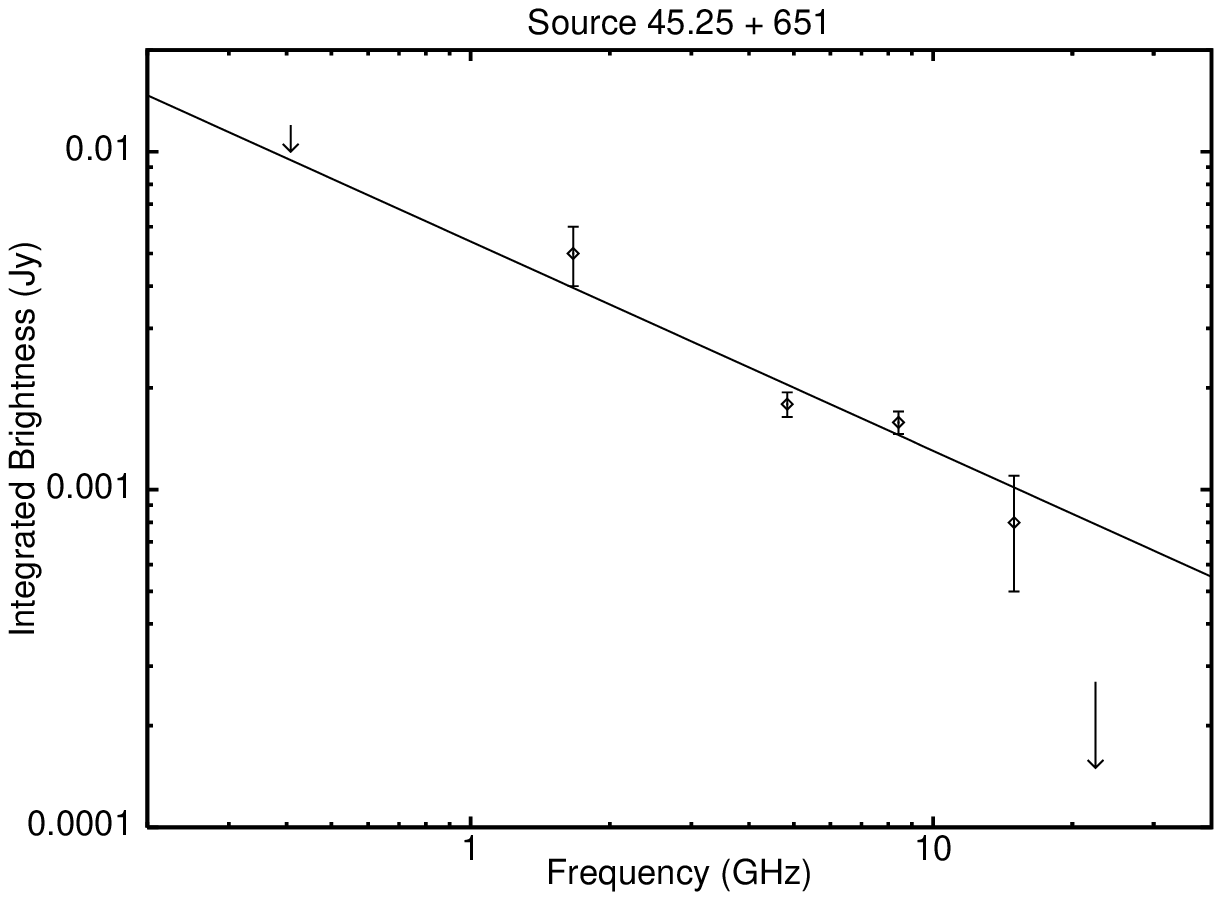}{0pt}{0}{70}{70}{-20}{-190}
\end{figure}

\clearpage

\begin{figure}
\figurenum{}
\plotfiddle{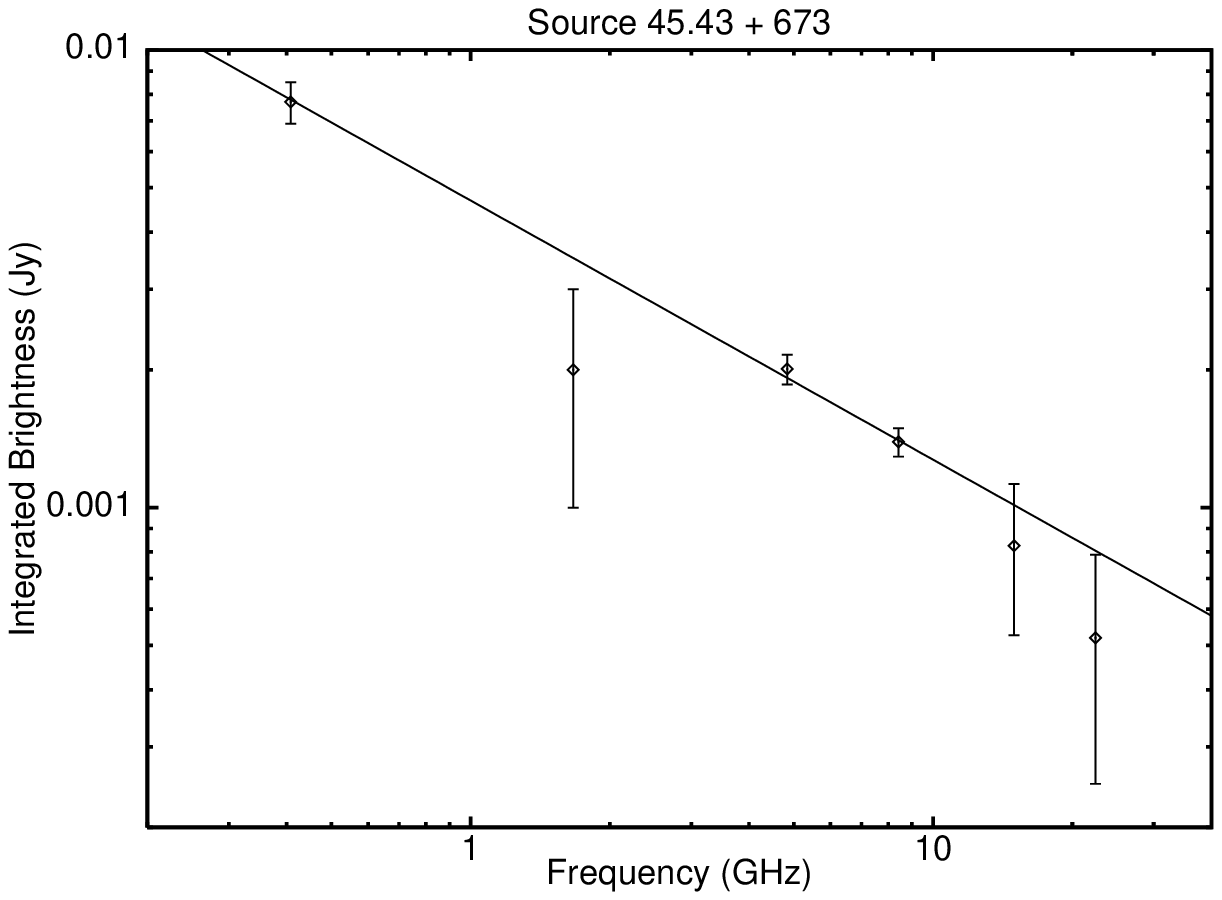}{0pt}{0}{70}{70}{-300}{-46}
\plotfiddle{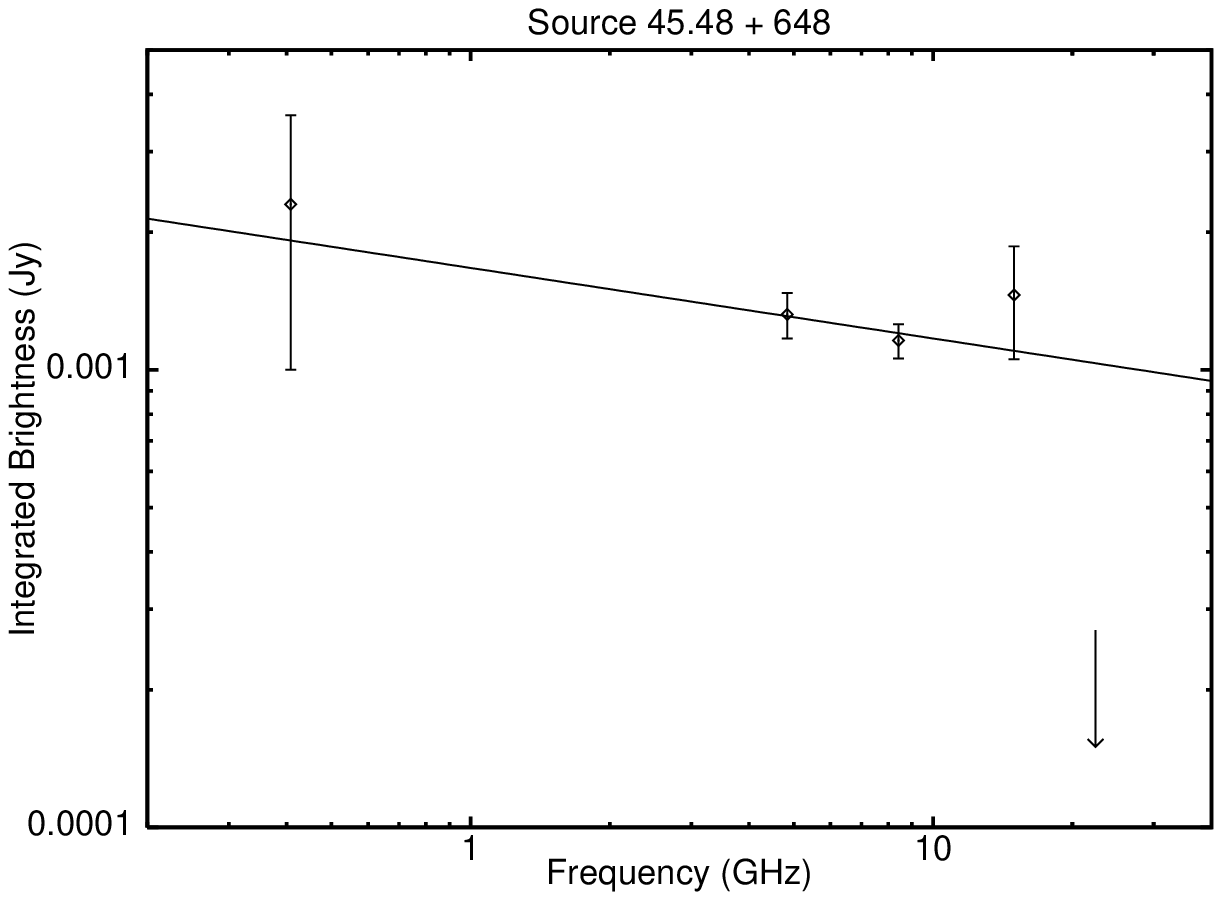}{0pt}{0}{70}{70}{-20}{0}
\plotfiddle{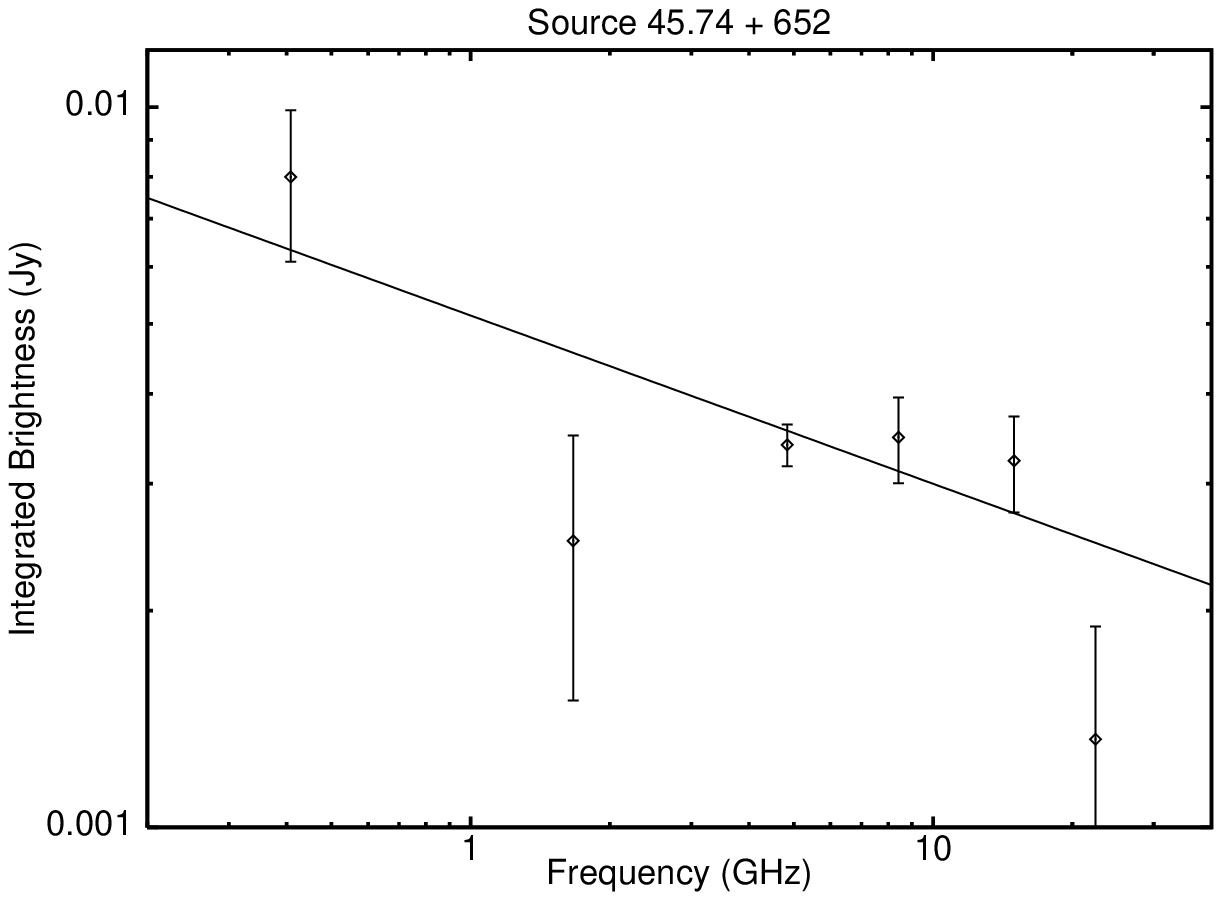}{0pt}{0}{70}{70}{-300}{-141}
\plotfiddle{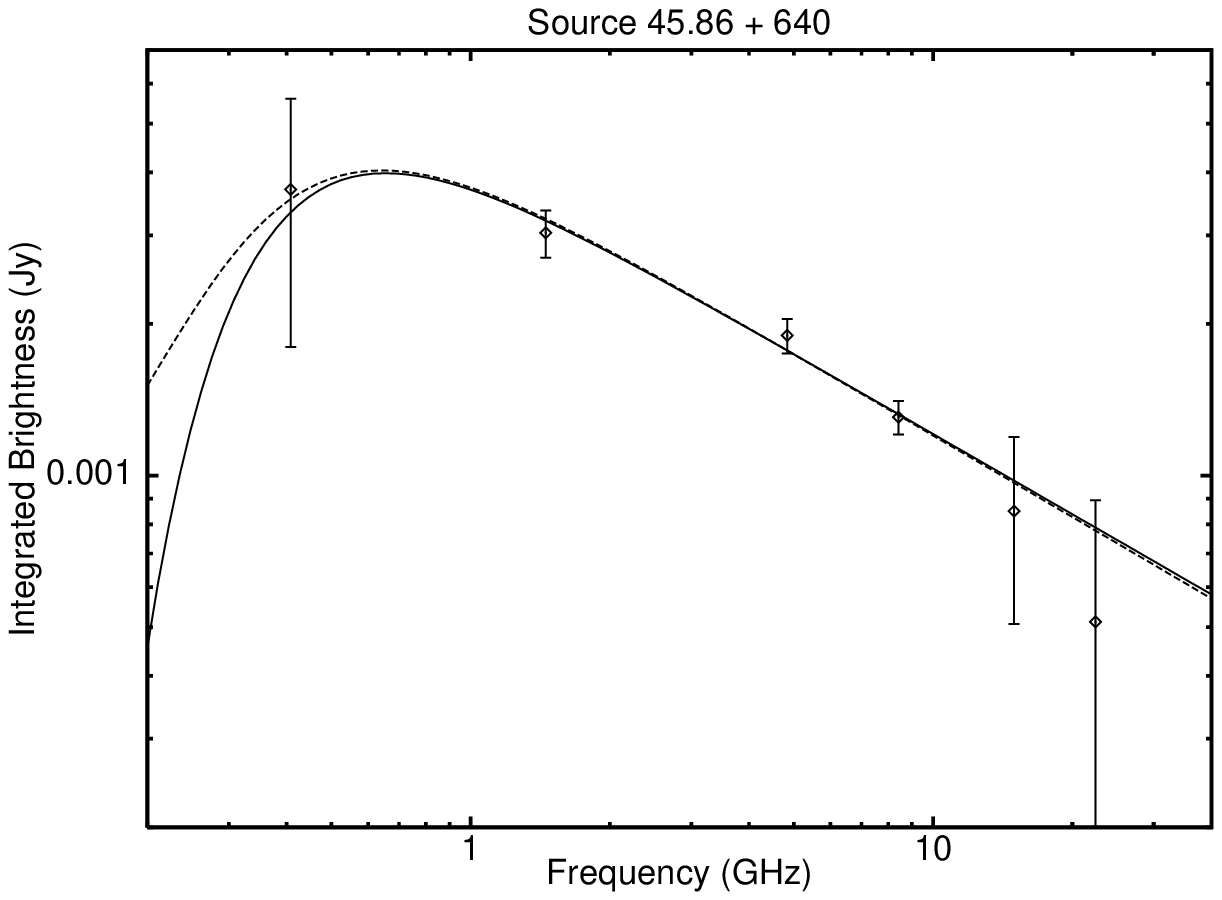}{0pt}{0}{70}{70}{-20}{-95}
\plotfiddle{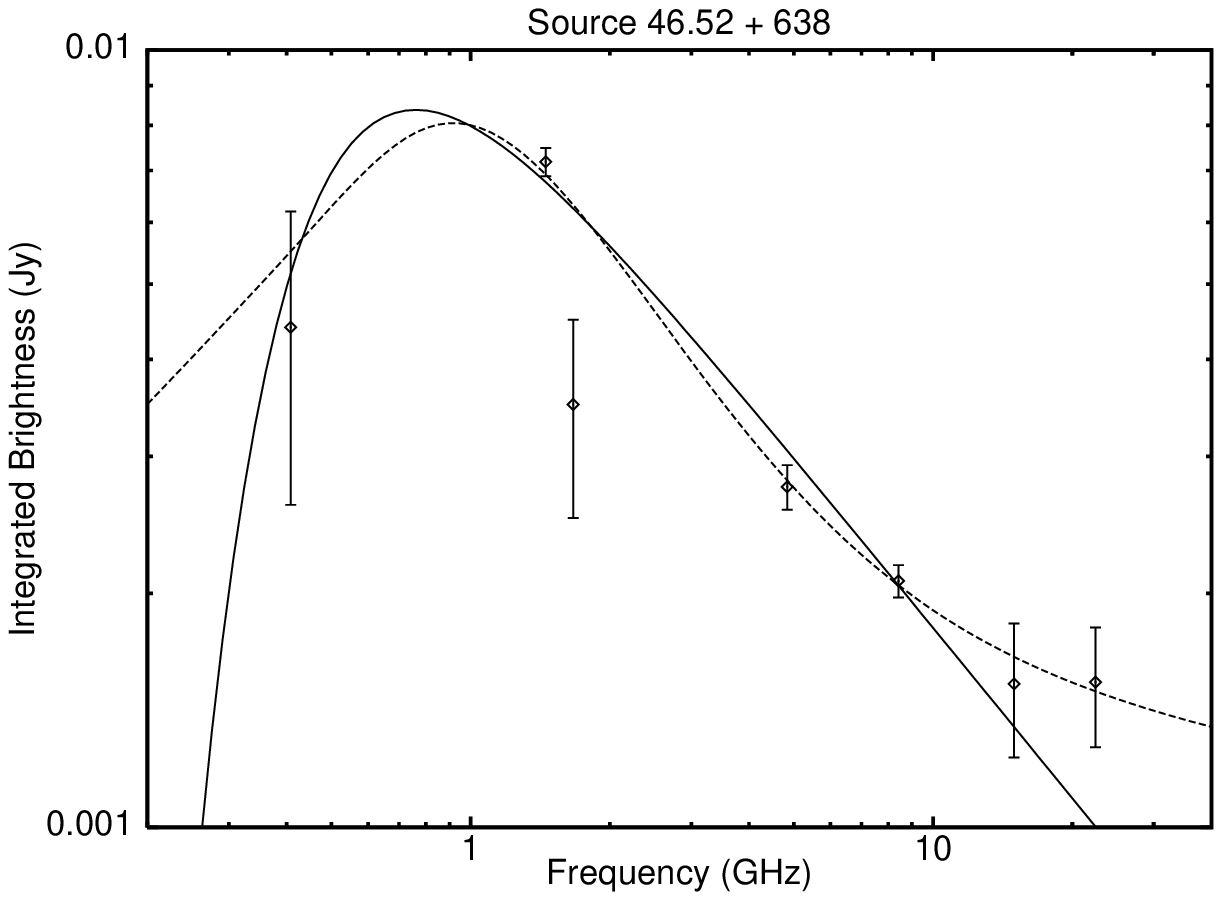}{0pt}{0}{70}{70}{-300}{-236}
\plotfiddle{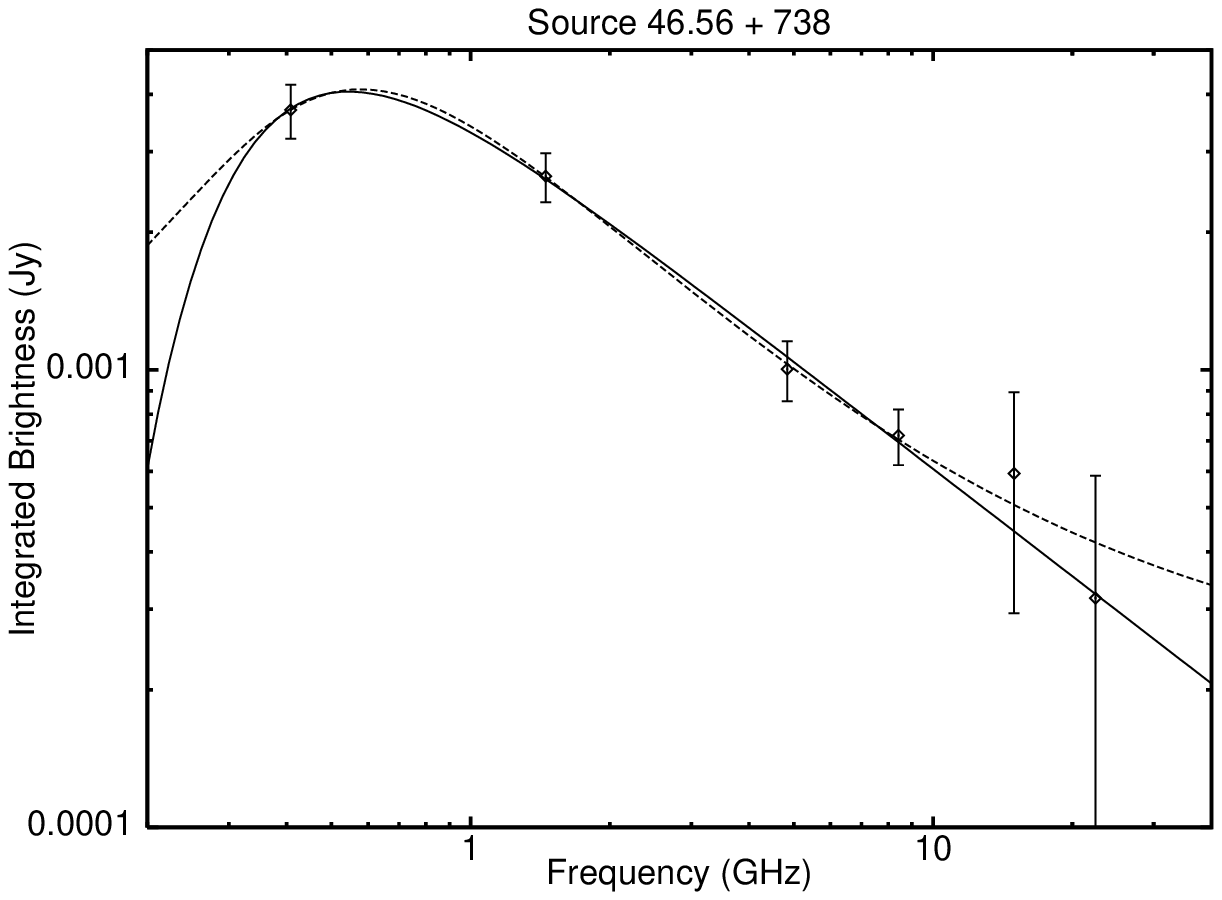}{0pt}{0}{70}{70}{-20}{-190}
\end{figure}

\clearpage

\begin{figure}
\figurenum{}
\plotfiddle{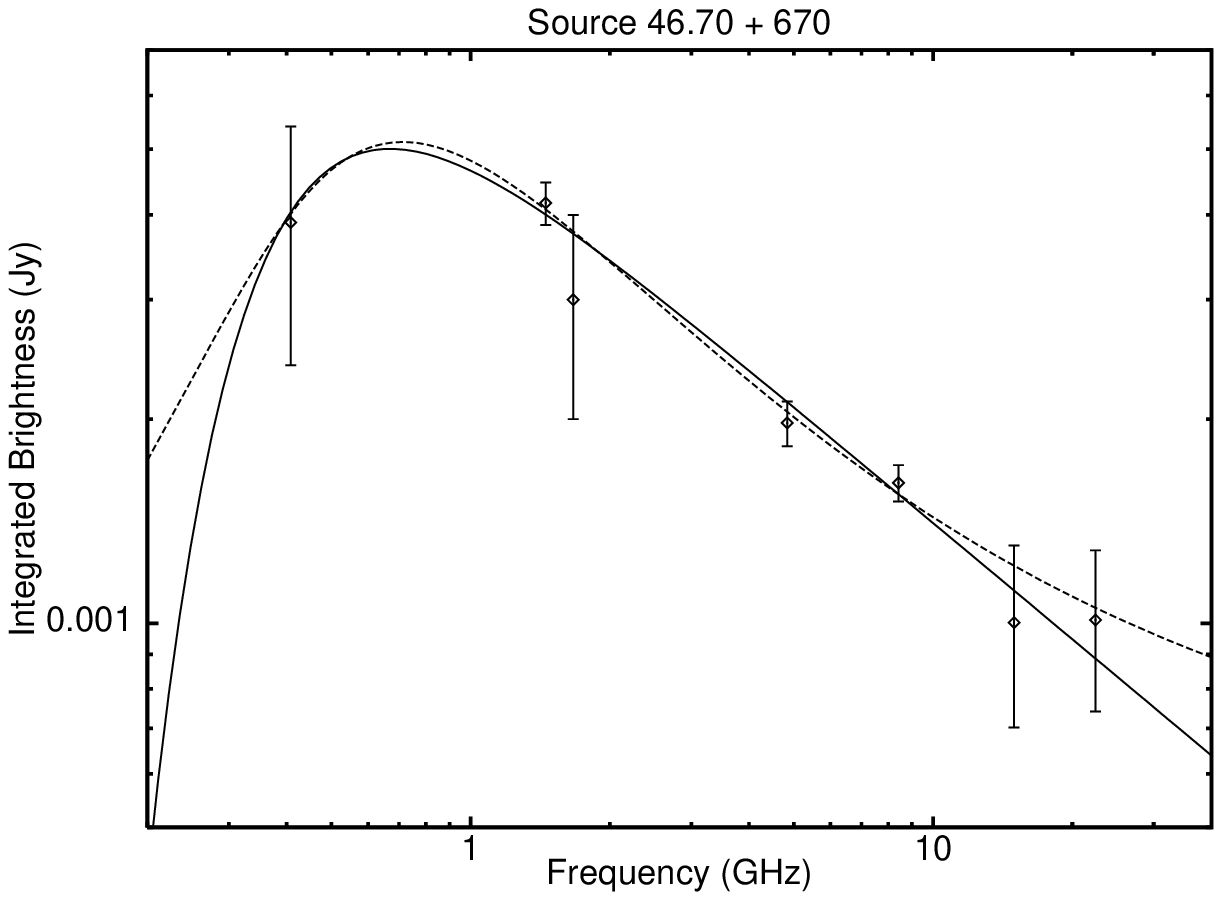}{0pt}{0}{70}{70}{-300}{-46}
\plotfiddle{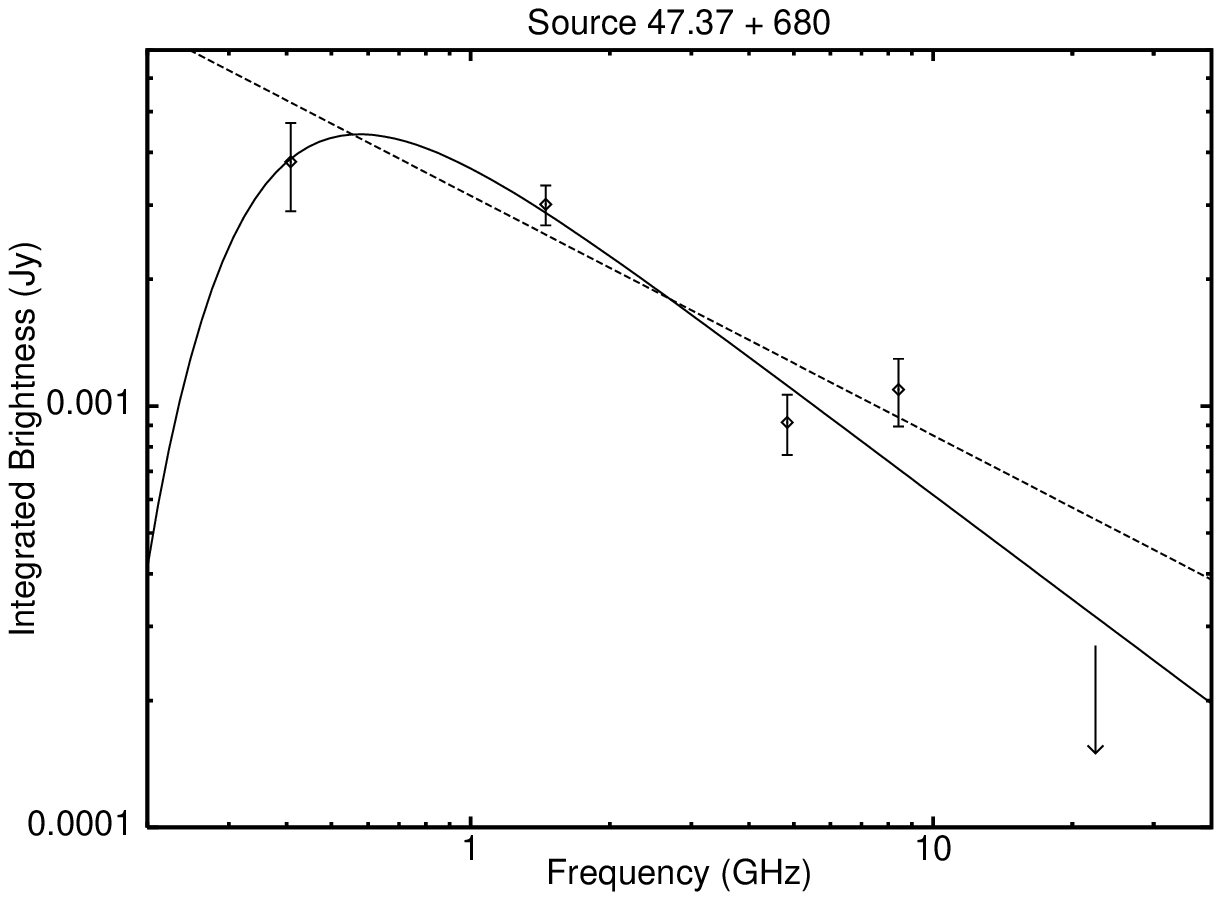}{0pt}{0}{70}{70}{-20}{0}
\end{figure}

\clearpage

%%FIGURE 2

\plotone{hist1.eps}

\clearpage

%%FIGURE 3

\plotone{hist2.eps}

\clearpage

%%FIGURE 4

\plotone{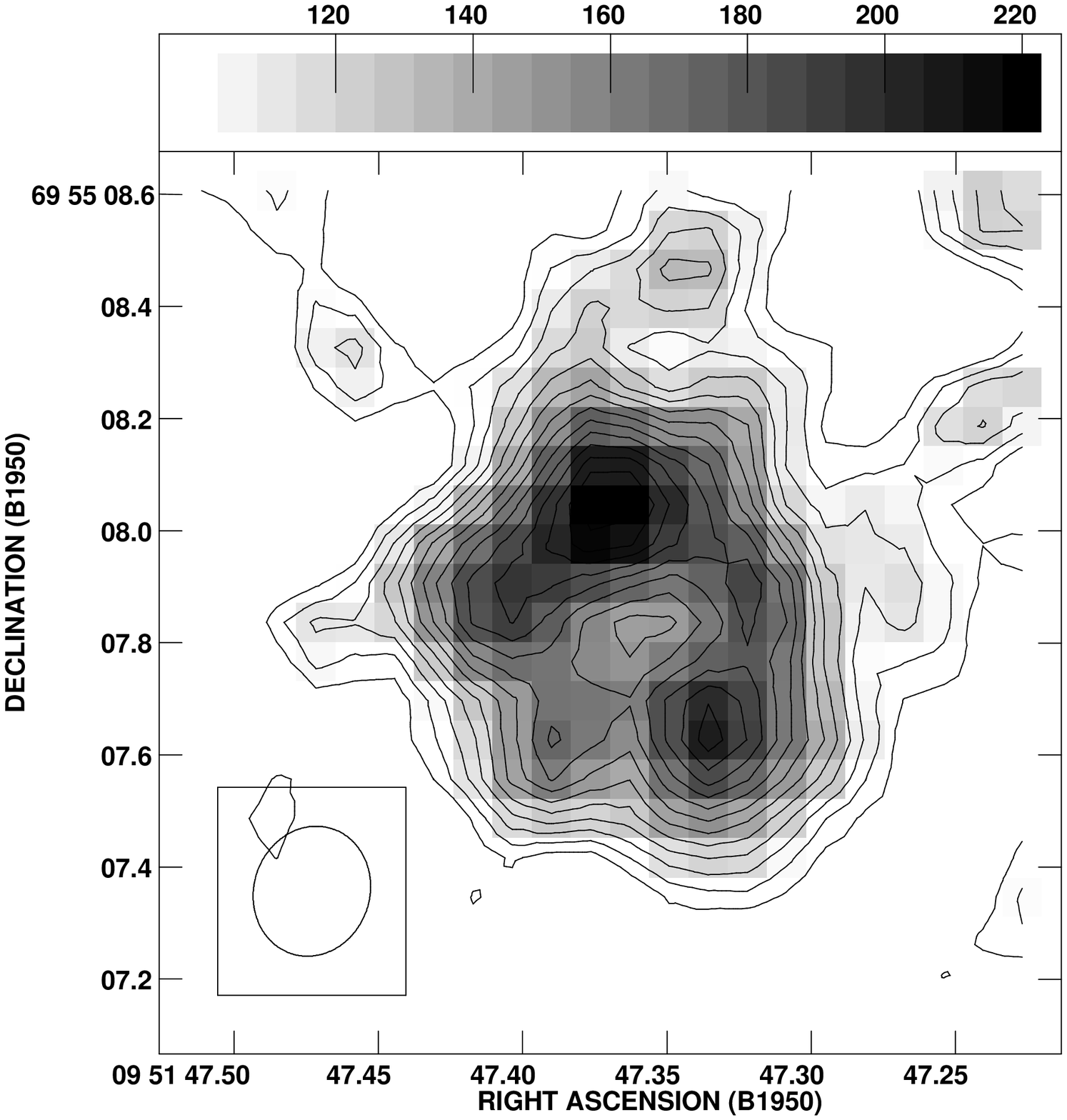}

\end{document}